\documentclass[twocolumn]{aastex631}
\usepackage{amsmath}

\usepackage{amsmath}

\begin{document}
\title{Repeated partial disruptions in a WD-NS or WD-BH merger modulate the prompt emission of long-duration merger-type GRBs}
%\footnote{Corresponding author: shenrf3@mail.sysu.edu.cn}

\author{Junping Chen} \affiliation{School of Physics and Astronomy, Sun Yat-Sen University, Zhuhai 519082, China; shenrf3@mail.sysu.edu.cn} \affiliation{CSST Science Center for the Guangdong-Hongkong-Macau Greater Bay Area, Sun Yat-Sen University, Zhuhai 519082, China}
\author{Rong-Feng Shen} \affiliation{School of Physics and Astronomy, Sun Yat-Sen University, Zhuhai 519082, China; shenrf3@mail.sysu.edu.cn} \affiliation{CSST Science Center for the Guangdong-Hongkong-Macau Greater Bay Area, Sun Yat-Sen University, Zhuhai 519082, China}
\author{Wen-Jun Tan} \affiliation{Key Laboratory of Particle Astrophysics, Institute of High Energy Physics, Chinese Academy of Sciences, Beijing 100049, China; xiongsl@ihep.ac.cn}
\author{Chen-Wei Wang} \affiliation{Key Laboratory of Particle Astrophysics, Institute of High Energy Physics, Chinese Academy of Sciences, Beijing 100049, China; xiongsl@ihep.ac.cn}
\author{Shao-Lin Xiong} \affiliation{Key Laboratory of Particle Astrophysics, Institute of High Energy Physics, Chinese Academy of Sciences, Beijing 100049, China; xiongsl@ihep.ac.cn}
\author{Run-Chao Chen} \affiliation{School of Astronomy and Space Science, Nanjing University, Nanjing 210093, China; bbzhang@nju.edu.cn}
\author{Bin-Bin Zhang} \affiliation{School of Astronomy and Space Science, Nanjing University, Nanjing 210093, China; bbzhang@nju.edu.cn}

\begin{abstract}
The progenitors of gamma-ray bursts (GRBs) have long been an unresolved issue. GRB 230307A stands out as an exceptionally bright event, belonging to the long-duration GRBs but also exhibiting a late emission component reminiscent of a kilonova. Together with the similar events GRBs 060614 and 211211A, they make up a new sub-group of GRBs with intriguing progenitors. If such long-duration merger-type GRBs originated from the coalescence of a white dwarf (WD) with a neutron star (NS) or a black hole (BH), as proposed in the recent literature, then the larger tidal disruption radius of the WD, together with a non-negligible residual orbital eccentricity, would make repeated partial tidal disruptions inevitable. This may modulate the mass accretion and jet launching process at the NS or BH, resulting in a quasi-periodic modulation (QPM) in the light curve of the GRB, on the orbital period. The detection of potential QPMs during the early episode of prompt emission of these three GRBs supports this scenario, and the relatively slow QPM ($>$ 1 s) suggests that the lighter object can not be a NS. We propose that the progenitor system of GRBs 230307A, 060614, and 211211A consist of a WD of mass 1.3 $M_\odot$, 0.9 $M_\odot$ and 1.4 $M_\odot$, respectively, and a NS (or BH). After several cycles of modulations, the WD is completely destructed, and the accretion of the remaining debris dominates the extended emission episode.
\end{abstract}
%%%%%%%%%%%%%%%%%%%%%%%%%%%%%%%%%%%%%%%%%
%%%%%%%%%%%%%%%%%%%%%%%%%%%%%%%%%%%%%%%%%
\section{Introduction}
\label{sec:intro}

Gamma-ray bursts (GRBs) are among the brightest transient events in the Universe, and their progenitors have been a subject of great interest to the astronomical community. 
Long GRBs (duration of prompt emission $\textgreater$ 2 s) have long been thought to be associated with core-collapse events of massive stars, while short GRBs tend to be associated with mergers of binary compact objects \citep{1993ApJ...413L.101K, 2006ARA&A..44..507W, 2014ARA&A..52...43B,2015PhR...561....1K, 2018pgrb.book.....Z}.

Intriguingly, a few long-duration bursts, such as GRB 060614 ($T_{90} \approx 102$ s) and 211211A ($T_{90} \approx 43$ s), caused a lot of trouble. Their prompt emissions are of long duration, with no associated supernovae observed, but are associated with candidate kilonova emission components, suggesting that they are a special type of GRB \citep{2006Natur.444.1053G, 2022Natur.612..223R}. In addition, GRBs 060614 (z $\sim$ 0.125) and 211211A (z $\sim$ 0.076) have large position offsets, 0.8 kpc and 8.19 kpc, respectively, from their host galaxies \citep{2006Natur.444.1053G, 2022Natur.612..223R, 2022Natur.612..232Y}. A merger of binary compact objects is one site of rapid neutron capture (r-process) nucleosynthesis, therefore it is a very likely origin for this particular GRB type \citep{2006Natur.444.1053G, 2006Natur.444.1044G, 2022Natur.612..223R, 2022Natur.612..228T, 2022Natur.612..232Y, 2022ApJ...936L..10Z, 2023ApJ...947L..21Z, 2023ApJ...954L..17Y}. However, the cause of their long durations is puzzling and still unknown. 

As a recently reported event, the peak flux and peak luminosity of GRB 230307A in the 0.1-2 MeV band are as high as $4.5 \times 10^{-4 }$ erg cm$^{-2}$ s$^{-1}$ and $4.9 \times 10^{51}$ erg s$^{-1}$, respectively, making it the secondly brightest GRBs (after GRB 221009A, \citeauthor{2023arXiv230705689S} \citeyear{2023arXiv230705689S}, \citeauthor{2023ApJ...954L..29D} \citeyear{2023ApJ...954L..29D}, \citeauthor{2023arXiv231007205Y} \citeyear{2023arXiv231007205Y}). The duration of its prompt emission exceeds two minutes, indicating that it belongs to Long GRBs. Subsequent observations suggest a strong association with a nearby galaxy at a redshift of 0.065, with offsets of up to 37 kpc from the center of the host galaxy \citep{2024Natur.626..737L, 2023arXiv230705689S}. 

Despite the long duration of the prompt emission, its association with a kilonovae feature (no supernovae was detected) suggests that the progenitor of GRB 230307A may be linked to a binary compact object system \citep{2024Natur.626..737L, 2024Natur.626..742Y, 2023arXiv230705689S}. Based on these properties, it can be classified together with GRBs 060614 and 211211A.

Several recent theoretical works have focused on the possible progenitors of GRB 230307A, and they all point to a merging binary compact object system
\citep{2024ApJ...963L..26Z, 2024ApJ...962L..27D, 2024Natur.626..742Y, 2024ApJ...962L..37D, 2024ApJ...964L...9W}.  In particular, \cite{2024ApJ...964L...9W} consider the merger of a white dwarf (WD)-neutron star (NS) binary, motivated by the need to explain the long emission duration. They propose that the kilonova-like signature may be powered by an ejecta heated by the magnetar spin-down energy output or the radioactive decay of $^{56}$Ni.

If this particular class of GRBs does indeed originate from a WD-NS or WD-black hole (BH) binary merger \citep{1999ApJ...520..650F, 2007MNRAS.374L..34K, 2018MNRAS.475L.101D, 2022Natur.612..232Y, 2023ApJ...947L..21Z, 2024ApJ...964L...9W}, given the relatively large tidal disruption radius of the WD and a probable residual orbital eccentricity, the repeated partial disruptions (RPDs) of the WD will be inevitable. This process would likely modulate the luminosity variation of an accretion-driven jet, thus leading to a quasi-periodic modulation (QPM) feature in the GRB light curve. This motivates us 1) to investigate the viability of this physical scenario, and 2) to look for possible QPM evidence in the prompt emission of these three long-duration merger-type GRBs.

Note that recently \cite{2024arXiv240812654L} suggest WD-BH merger as the progenitors of low-redshift long GRBs, as a solution to the puzzling deviation of their rates from the cosmic star formation rate history.

The paper is organized as follows. Section \ref{sec2} presents the model. Section \ref{sec3} shows our searches of QPM in the prompt emission light curves of GRBs 230307A, 060614 and 211211A, and Section \ref{sec3_1} presents the application of our model to these findings. Conclusions and discussion are given in Section \ref{sec4}.
%%%%%%%%%%%%%%%%%%%%%%%%%%%%%%%%%%%%%%%%%
%%%%%%%%%%%%%%%%%%%%%%%%%%%%%%%%%%%%%%%%%
\section{Repeated Partial Disruptions of WD}
\label{sec2}
In the merging binary compact object scenario, the binary loses orbital energy and angular momentum due to gravitational wave (GW) radiation, resulting in a continuous decrease of the orbital semimajor axis $a$ and the pericenter radius $R_p \equiv a(1-e)$ where $e$ is the orbital eccentricity. It can be shown that $a$ and $e$ are related as $a=c_0e^{12/19}[1+(121/304)e^2]^{870/2299}/(1-e^2)$ \citep{1964PhRv..136.1224P}, so that the orbital is being circularized as it shrinks. In a NS-NS or NS-BH system, $a$ eventually approaches the size of the NS $\sim$ 10 km, where $e \simeq 0$.

However, in a WD-NS or WD-BH binary, when $R_p$ reaches the Roche limit radius, the WD starts to be tidally stripped, at which point $a$ is much larger, and $e$ is small but non-negligible. Partial tidal disruptions may occur a number of times, on the orbital period P, which we call RPDs. The accretion of the stripped stellar material would be modulated on the same period. 

A few versions of this scenario have been proposed for various types of luminous quasi-periodic outbursting sources \citep{2019ApJ...871L..17S, 2022MNRAS.515.4344K, 2022ApJ...941...24K, 2023ApJ...944..184L}. Here we consider the possibility that, in the case of a relativistic jet being generated, it may give rise to a QPM in the GRB prompt emission.

%%%%%%%%%%%%%%%%%%%%%%%%%%%%%%%%%%%%%%%%%
\subsection{Period of the QPM}\label{sec2_1}
The radius of a WD of mass $M_*$ is given by
\citep{1983ApJ...267..315P}, 
\begin{equation}
R_*=9\times10^8\left[1-\left(\frac{M_*}{M_{ch}}\right)^{4/3}\right]^{1/2}(m_*)^{-1/3} \ {\rm cm},
\label{eq:LebsequeIp1} \end{equation}
where $M_{ch}$ = 1.44 $M_\odot$ is the Chandrasekhar mass and $m_*=M_*/M_\odot$. 
Consider that the WD has a more massive compact companion whose mass is $M\geq M_*$. As the binary orbit shrinks and $R_p$ reaches twice the tidal disruption radius \citep{1971ARA&A...9..183P, 1988Natur.333..523R}
\begin{equation}
R_t=R_*\left(\frac{M+M_*}{M_*}\right)^{1/3},
\label{eq:LebsequeIp2} \end{equation}
the material in the surface layers of the WD begins to be stripped and move toward the compact companion.

Using $R_p \equiv (1-e)a \backsimeq 2R_t$ and $a^3/P^2=G(M+M_*) / 4\pi^2$ with Eq. \ref{eq:LebsequeIp2}, one would know that when RPD occurs, the orbital period is 
\begin{equation}
\begin{split}
P= 2\pi \sqrt {\frac{(2R_*)^3}{GM_*(1-e)^3}} \\ \backsimeq 4.6 \left( \frac{R_*} {3 \times 10^{-3} R_\odot} \right)^{3/2} m_*^{-1/2} (1-e)^{-3/2} \ {\rm s},
\label{eq4}     
\end{split}
\end{equation}
independent of $M$. The central compact object could be a NS or a BH. If it is a BH, it can not be too massive, otherwise, it will directly engulf the WD. This requires that $R_s \equiv 2GM /c^2< R_t$, which leads to a loose constraint $ M <$  10$^6$ $M_\odot$.

The stripped material would form a small accretion disk around the NS / BH. As the WD's orbital pericenter slowly moves in, the stripped mass per orbit grows, and eventually a relativistic jet may be launched from the disk. Due to the residual orbital $e$, the stripping occurs only at $R_p$. This regular non-continuity, or periodic enhancement, of the matter supply to the accretion disk modulates the accretion process and thus the (kinetic) luminosity of the jet. Therefore, one should expect a QPM with a period of $P$, in the GRB light curve.

%%%%%%%%%%%%%%%%%%%%%%%%%%%%%%%%%%%%%%%%%
\subsection{GRB Jet Production and Energetics}\label{sec2_2}
How does a WD-NS/BH merger launch a GRB-producing jet? We first consider the case of the central compact companion being a NS. 
1) The jet is driven through an accretion disk \citep{2007MNRAS.374L..34K, 2019EPJA...55..132F}.  The energy driving the jet comes from the gravitational energy of the accreted material. $E_{disk}=GMm_{acc}/R_{NS} \simeq  3.7\times 10^{51} (M/1.4M_{\odot})(R_{NS}/10 ^{6}cm)(m_{acc}/0.01M_\odot)$ ergs, where $R_{NS}$ is the radius of the NS. If the energy conversion efficiency is $\eta$=3\%, then an accreted mass of $m_{acc}=0.1M_\odot$ is enough to drive a $10^{51}$ ergs explosion \citep{2019EPJA...55..132F}.
2) The jet may also be produced during WD-NS mergers when the toroidal magnetic field of the NS is amplified, triggering magnetic bubble eruptions \citep{1998ApJ...505L.113K, 2000ApJ...542..243R, 2021NatAs...5..911Z, 2022Natur.612..232Y, 2023ApJ...947L..21Z}. The luminosity of GRBs in this mechanism is related to the accretion rate as $L_{jet}=\eta \dot{M} c^2 \simeq  1.8\times 10^{51} (\eta /10\% ) (\dot{M}/0.01M_\odot s^{-1}) \ {\rm erg  \ s^{-1}}$.

For the case of a BH as the central compact companion, an accretion disk is essential in generating a jet, via either the $\nu-\bar{\nu}$ annihilation or the Blandford-Znajek mechanism. 
1) The luminosity released by neutrino annihilation can be simply estimated as \citep{1999ApJ...518..356P, 1999ApJ...520..650F} $log L_{\nu, \dot{\nu}} \simeq  44+5\times log(\dot{M}/0.01M_\odot s^{-1}) +3.4(J_{BH}c/GM^2) \ {\rm erg  \ s^{-1}}$, where $J_{BH}$ is the BH angular momentum. An accretion rate of $\sim 0.01 M_\odot s^{-1}$ typical for WD-BH mergers, with a beaming factor of 100, can explain bursts with isotropic-equivalent energies of $10^{48-51}$ ergs \citep{1999ApJ...520..650F}. However, it may be difficult to produce a GRB as energetic as GRB 230307A ($E_{iso}\sim 3 \times 10^{52} $ ergs).  
2) In a BH accretion disk, the magnetic field with a strength $B$ is stretched, twisted and enhanced during the accretion of matter, which extracts the rotational energy of the BH to produce the jet, $L_{BZ}=10^{50} (J_{BH}c/GM)^2 (M/3M_{\odot})^2(B/10^{15} G)^2 \ {\rm erg  \ s^{-1}}$ \citep{1977MNRAS.179..433B}. In addition, the energy driving the jets may also derive from the kinetic energy of the accretion disk material, same as we discussed in the case of NS accretion disks.

In either case, the property of the inner region of the accretion disk is directly related to the GRBs. Therefore, the disk formation and evolution during the WD-NS/BH merger have been extensively studied \citep{1984MNRAS.208..721P, 1989ApJ...346..847C, 1999ApJ...520..650F, 2012MNRAS.419..827M, 2013ApJ...763..108F, 2016MNRAS.461.1154M, 2019MNRAS.486.1805Z, 2020MNRAS.493.3956Z, 2021MNRAS.506.3511M, 2022MNRAS.510.3758B, 2023ApJ...956...71K, 2024A&A...681A..41M}. 
In particular, \cite{2012MNRAS.419..827M} proposed that the heating rate of the inner disk ($r\lesssim 10^{8.5-9}$ cm) during their early evolution was significantly higher due to thermonuclear burning. \cite{2013ApJ...763..108F} subsequently show that nuclear burning indeed affects the dynamics of the accretion flow, and may imprint variability (on time scale $\sim$ seconds) in the GRB emission if nuclear burning causes large amplitude fluctuations in the central accretion rate. However in our model, the material transfer in the RPD process occurs only near $R_p$,  and significantly more material is stripped away (see Section \ref{sec2_3}) when relativistic jets and GRBs are generated. Therefore, this QPM effect is much stronger. Even though nuclear burning might affect the accretion flow dynamics, it would still be hard to mask the QPM in the GRB light curve.

%%%%%%%%%%%%%%%%%%%%%%%%%%%%%%%%%%%%%%%%%
\subsection{Duration of the QPM Episode}\label{sec2_3}
In this scenario, what sets the duration of the prompt emission of GRB? Firstly, one could consider the GW-driven orbital shrinkage time scale $t_{GW}(a) \equiv a /|\dot{a}| \equiv E/ |\dot{E} |$, where the binary orbital energy $E=-G M_*(M_*+M)/(2a)$ decreases at a rate of \citep{1983bhwd.book.....S, 1963PhRv..131..435P} 
\begin{equation}
\dot{E}=-\frac{32}{5} \frac{G^4}{c^5} \frac{(M_*+M)^3 M_*}{a^5} \frac{f(e)}{(1-e^2)^{7/2}},
\label{eq:LebsequeIp10} \end{equation}
and $f(e)=1 +\frac{73}{24}e^2 +\frac{37}{96}e^4$. So
\begin{equation}
\begin{split}
t_{GW}\backsimeq 8.4 \times 10^4 m_*^{-1} \left( \frac{P}{1 \ {\rm s}}\right)^{8/3} \left( \frac{M} {10 M_\odot} \right)^{-2/3}  \frac{(1-e^2)^{7/2}}{f(e)} \ {\rm s}. 
\end{split} \label{eq:LebsequeIp11} \end{equation}
This time scale is too long compared with the prompt durations of the long-duration merger-type GRBs. Considering $1-e < 1$, the factor $(1-e^2)^{7/2} / f(e)$ could shorten $t_{GW}$ but not by much.

However, in the actual physical scenario, when the RPD process first occurs, the WD is stripped of too little material to form a hyper-accreting disk and to generate a relativistic jet. Only when the stripped mass is significant enough, it is possible to produce a GRB. Furthermore, the WD would inevitably undergo a volume expansion when the stripped mass is substantial. This expansion makes it even more susceptible to further partial disruptions, resulting in an avalanche-like process that brings the RPD process to a quick end. These two factors might substantially shorten the GRB-producing RPD. Below we estimate this duration.

\begin{figure}
\centering
\includegraphics[height=5.1cm,width=7.4cm]{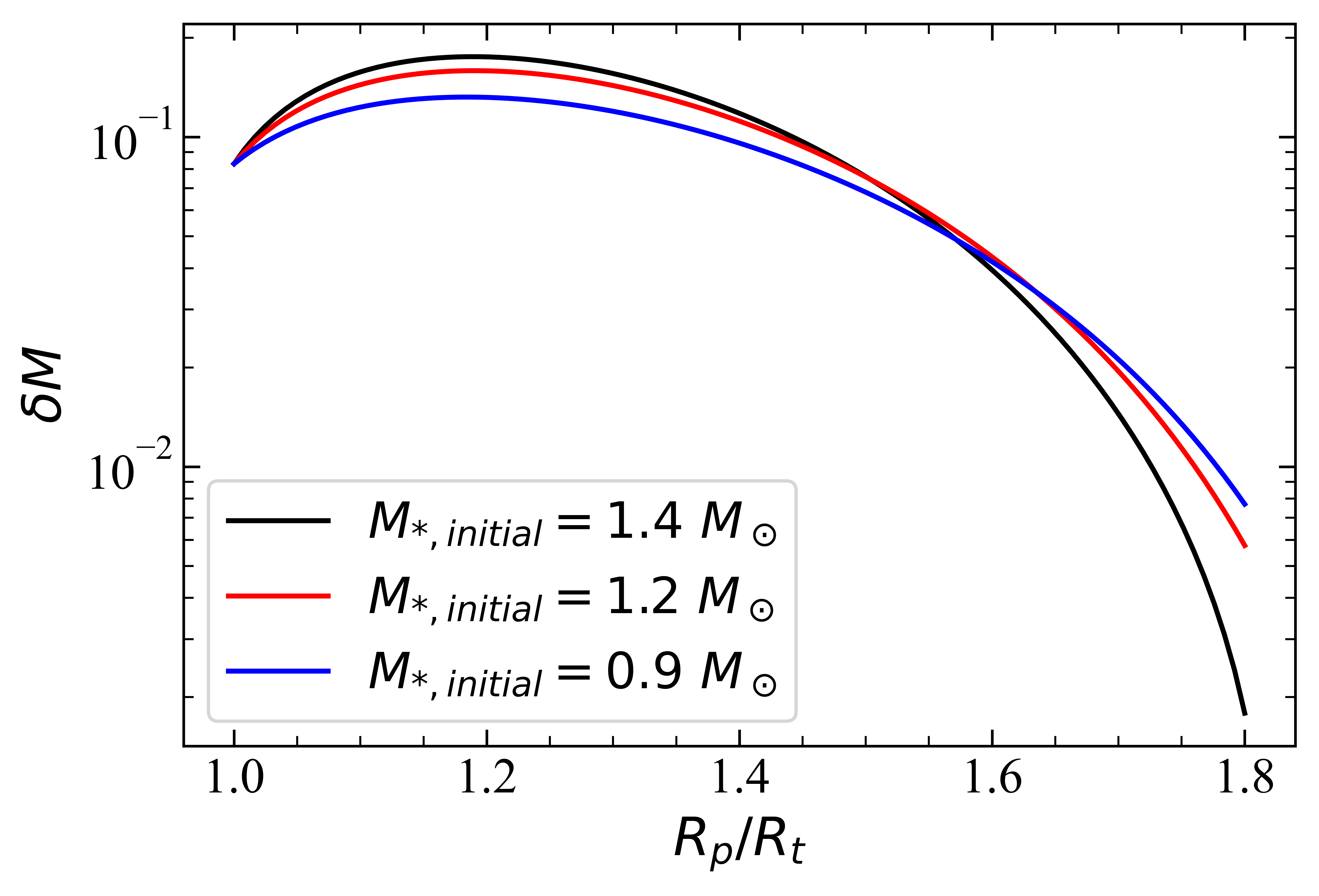}
\caption{The evolution of the stripped mass $\delta M$ with ever decreasing $R_p/R_t$ during the RPD process, calculated from Eqs. (\ref{eq:LebsequeIp1}-\ref{eq:LebsequeIp2}) and (\ref{eq:LebsequeIp13}). The black, red and blue lines are for different initial WD masses. Note that we start the process all initially at $R_p/R_t$=1.8 and later keep $R_p$ fixed. As the RPD proceeds, $M_*$ decreases due to mass stripping, thus $R_t$ increases.}
\label{Fig2_1}
\end{figure}

As $R_p$ approaches 2$R_t$, the WD begins to lose mass near the pericenter, where the WD transiently spills out of the Roche lobe ($R_*>R_{lobe}$), $R_{lobe} \backsimeq 0.5 R_p (M/M_*)^{-1/3}$ (e.g., \citeauthor{1971ARA&A...9..183P} \citeyear{1971ARA&A...9..183P}), and the surface material of the WD is stripped away \citep{2022ApJ...941...24K}. Assuming that the stripped layer is a spherical shell with a surface thickness of $z=R_*-R_{lobe}$, then the stripped mass is 
\begin{equation}
\delta M =4 \pi R_*^2 \int^{z}_{0} \rho (z') dz',
\label{eq:LebsequeIp12} 
\end{equation}
where the density profile at the WD surface is approximated as $\rho (z') \backsimeq 2.3 \times 10^{-7} m_*^{3/2} (10^{-2}R_\odot/R_*)^{3}(z')^{3/2} \ {\rm g} \ {\rm cm^{-3}}$ \citep{2023ApJ...947...32C}. Thus, the stripped mass is obtained as
\begin{equation} 
\frac{\delta M}{M_*} \backsimeq 4.8\left[1-\left(\frac {m_*}{1.44}\right)^{4/3}\right]^{3/4} \left(1-\frac{R_p}{2R_t}\right)^{5/2}.
\label{eq:LebsequeIp13} 
\end{equation}
As $M_*$ decreases and thus $R_t$ increases, $\delta M / M_*$ increases significantly (as shown in Figure \ref{Fig2_1}). 
\begin{figure}
\centering
\includegraphics[height=4.3cm,width=8.4cm]{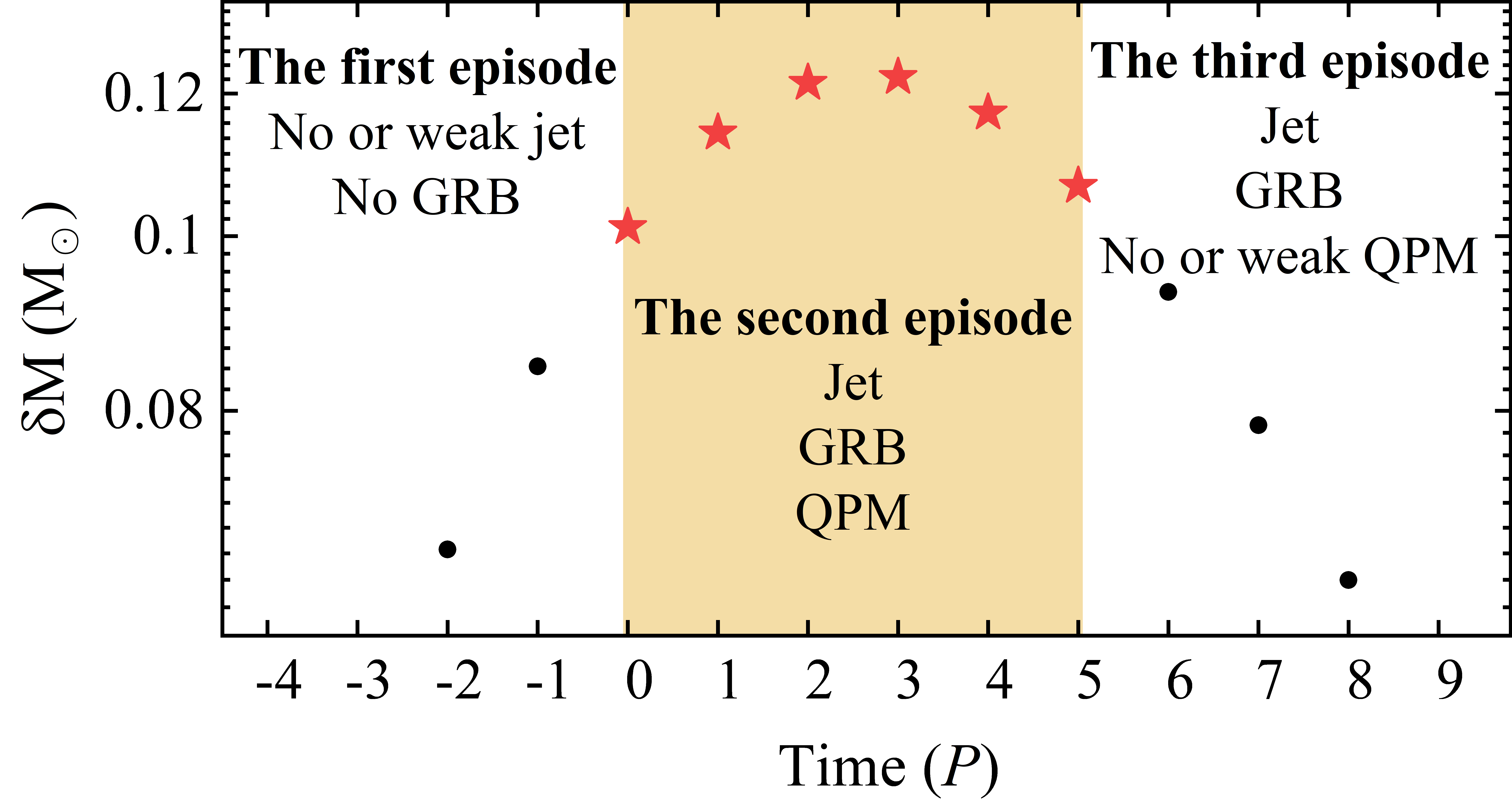}
\caption{The evolution of the stripped mass with time (in units of orbital period $P$) for a WD with the initial mass of 1.3$M_\odot$, as calculated by Eqs. (\ref{eq:LebsequeIp1}-\ref{eq:LebsequeIp2}) and (\ref{eq:LebsequeIp13}). The zero time here is chosen to be when the stripped mass firstly reaches a GRB-jet generating threshold, which is set to be $\delta M \simeq 0.1M_\odot$. The first episode lasts very long ($\sim 10^4$ orbits, see \citeauthor{2023ApJ...947...32C} \citeyear{2023ApJ...947...32C}), and the second episode produces GRB and shows QPM, while the third episode has no or weak QPM and the RPD process ends at this episode.}
\label{Fig2_2}
\end{figure}

The RPD actually begins when $R_p=2R_t$, but at this point, too little material is stripped off. The accretion of this material by the compact companion is not strong enough to produce a relativistic jet, or the jet would be too weak to be detectable. As $R_t$ slowly expands and reaches $R_p$, $\delta M$ increases rapidly and once it reaches a threshold, the GRB jet is generated and the QPM may be observed in the GRB light curve. Since the $\delta M / M_*$ is relatively large at this time, the WD may be completely destroyed within a few $P$. 

\begin{figure*}
\centering
\includegraphics[height=3.5cm,width=7.8cm]{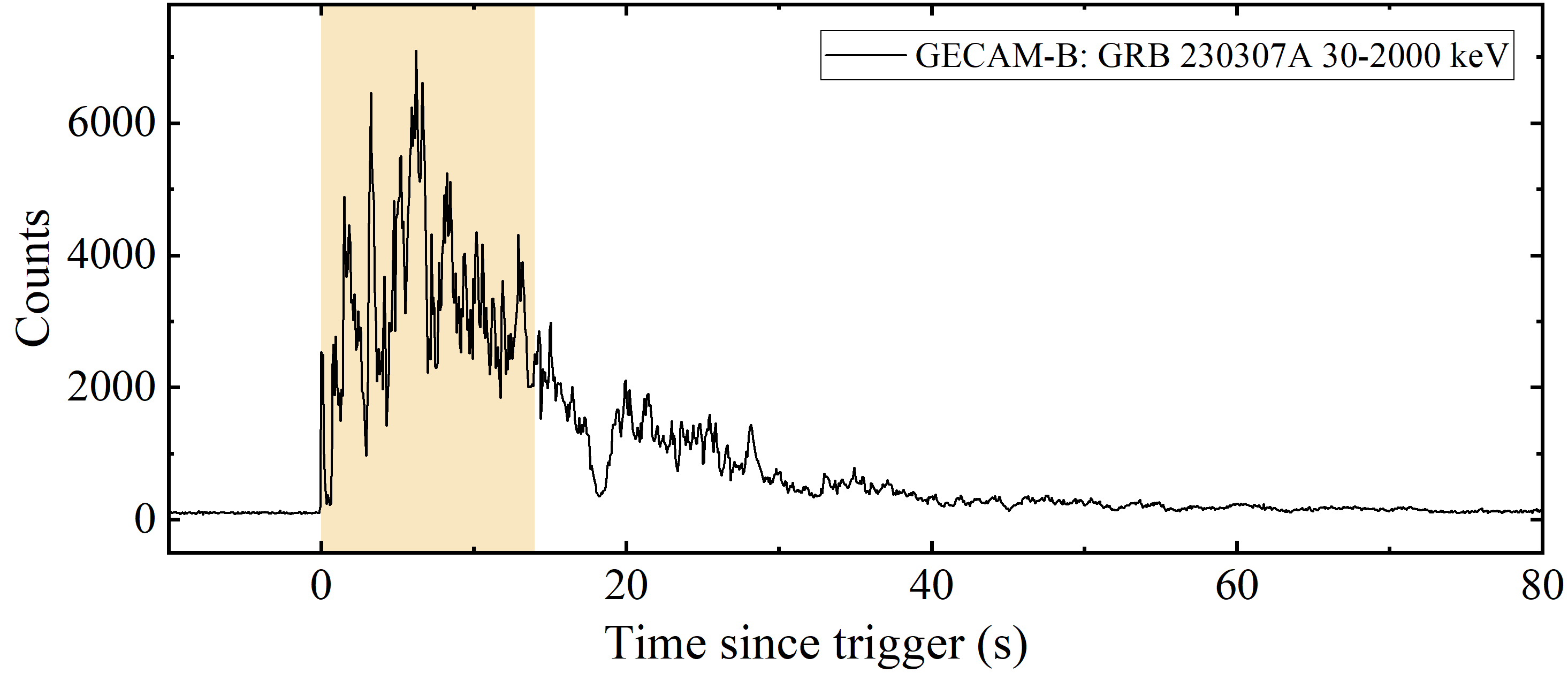}
\centering
\includegraphics[height=4.4cm,width=7.8cm]{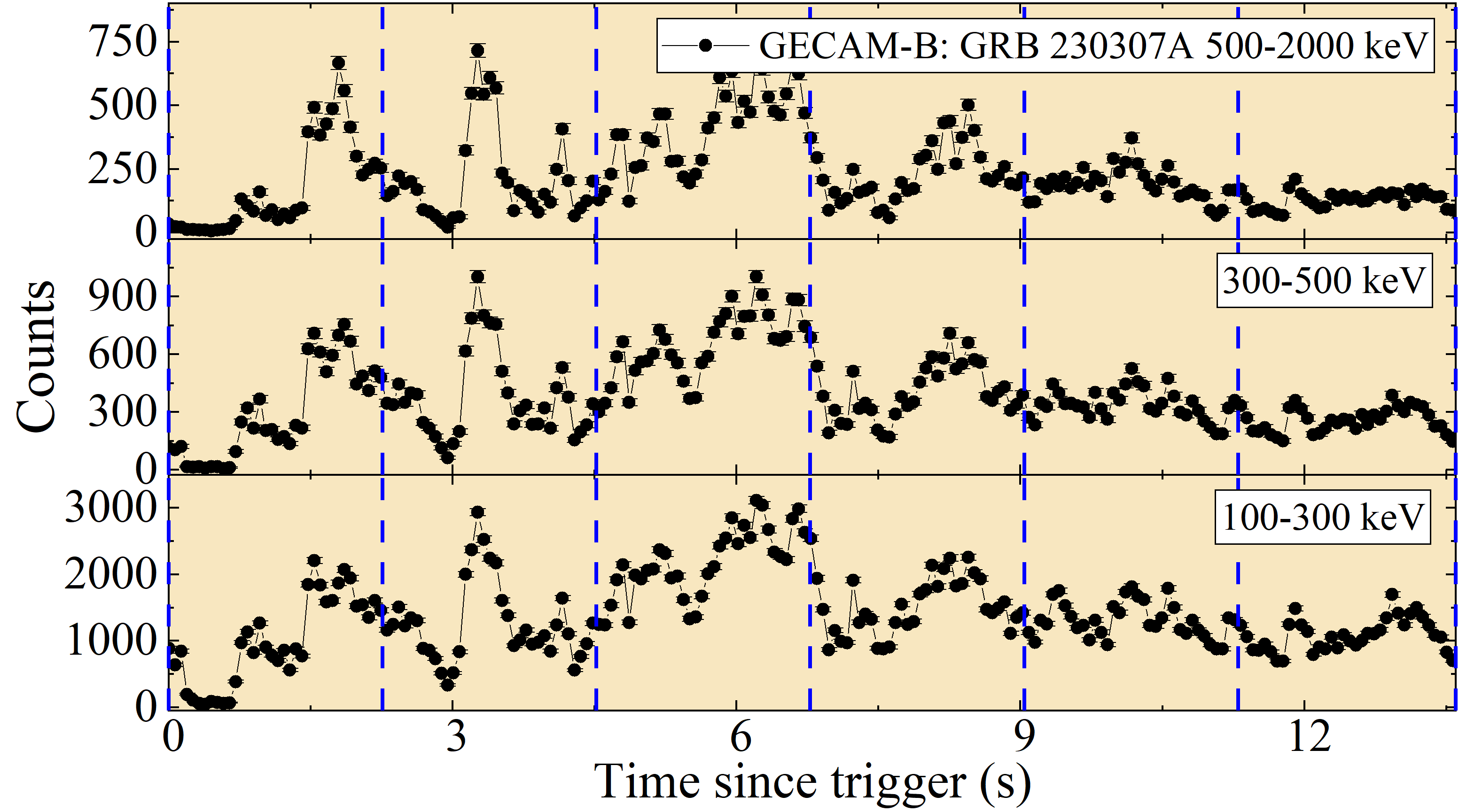}
\caption{The left panel presents the light curve of GRB 230307A in the 30-2000 keV energy range, where the light yellow region indicates the QPM episode. The right panel shows an enlarged view of the QPM episode, in three energy bands, respectively. The blue dashed lines indicate the cycles of the QPM.}
\label{Fig3_1}
\end{figure*}
\begin{figure*}
\centering
\includegraphics[height=3.6cm,width=7.8cm]{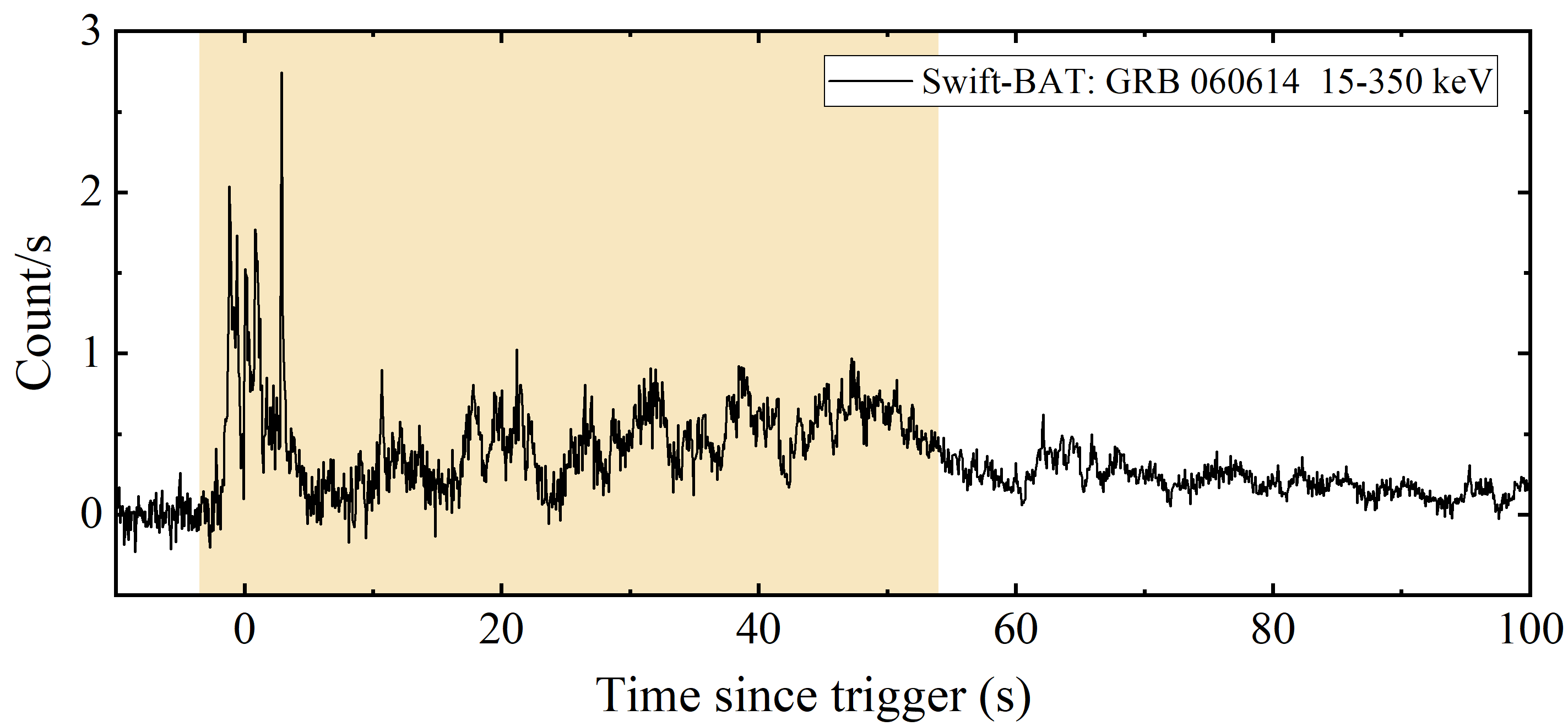}
\centering
\includegraphics[height=3.5cm,width=7.8cm]{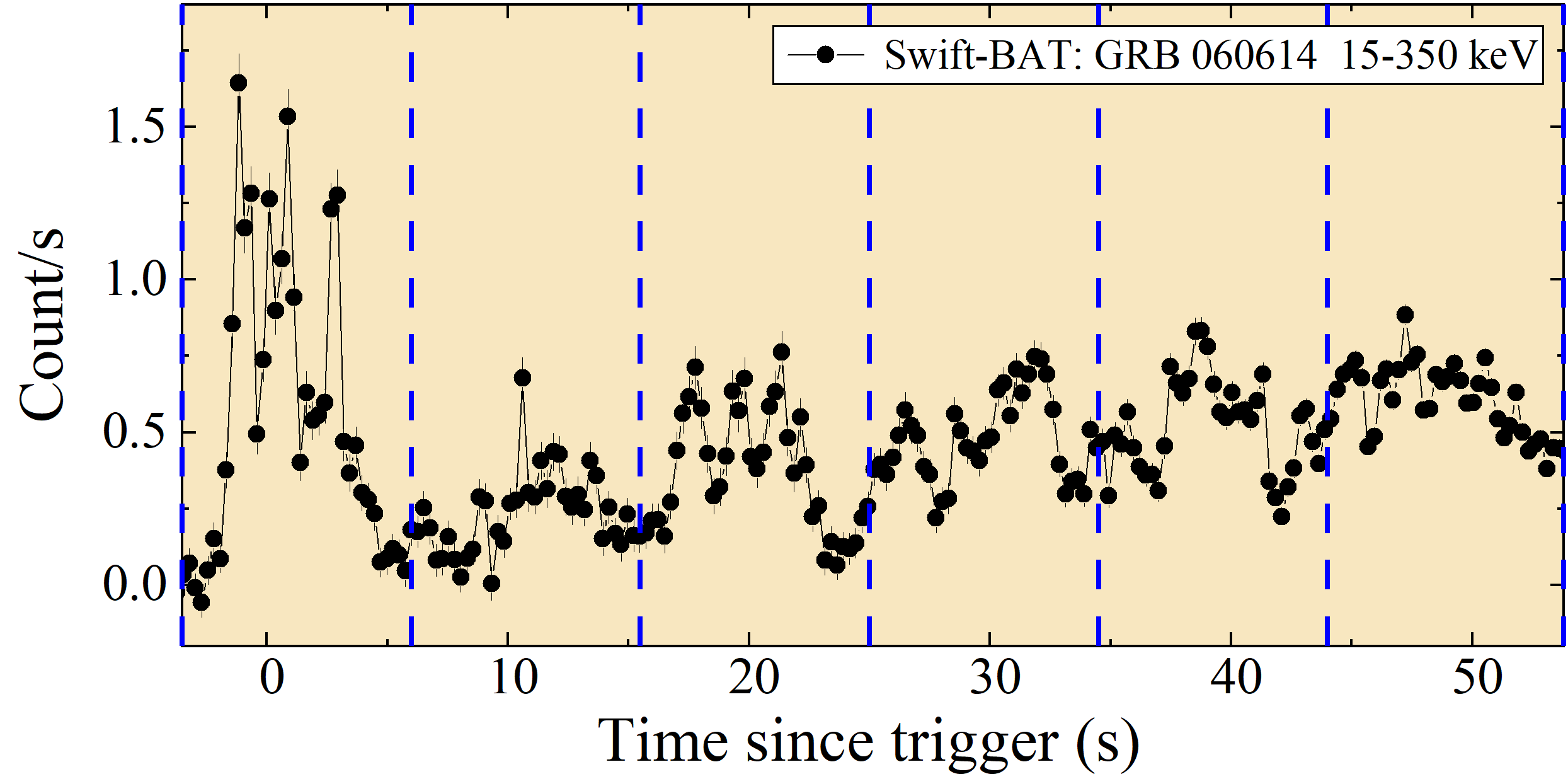}
\caption{Same as in Figure \ref{Fig3_1} but for GRB 060614 (15-350 keV).}
\label{Fig3_2}
\end{figure*}
\begin{figure*}
\centering
\includegraphics[height=3.3cm,width=7.8cm]{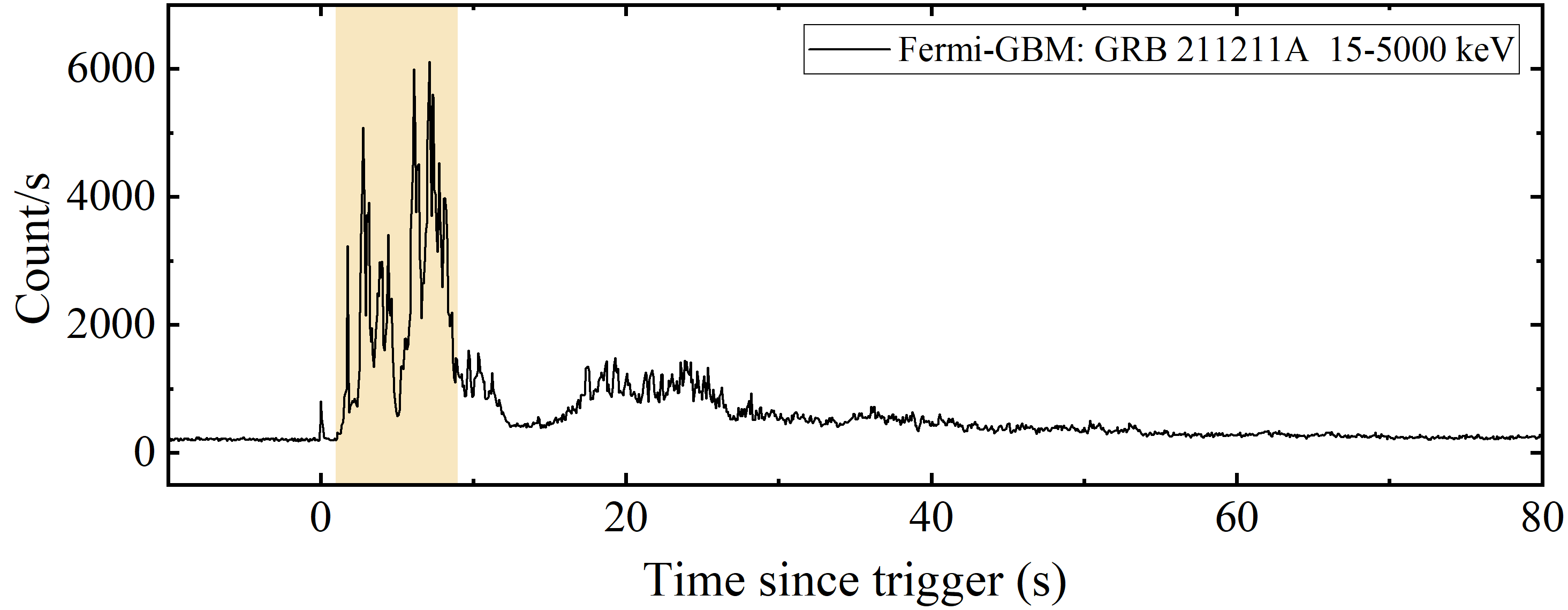}
\centering
\includegraphics[height=3.3cm,width=7.8cm]{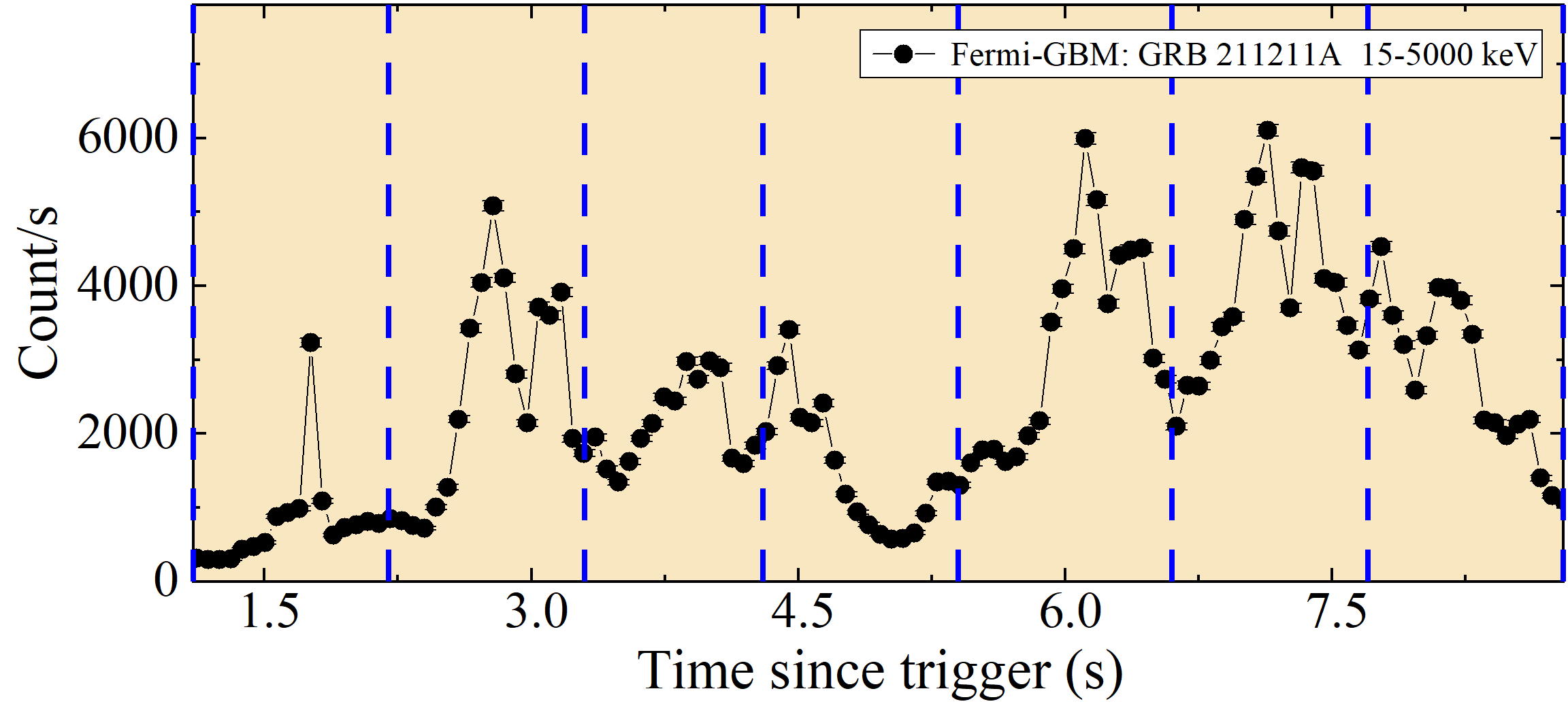}
\caption{Same as in Figure \ref{Fig3_1} but for GRB 211211A (15-5000 keV).}
\label{Fig3_3}
\end{figure*}

As an example, Figure \ref{Fig2_2} shows the evolution of the stripped mass for a 1.3 $M_\odot$ WD (this mass is chosen to match the case for GRB 230307A; see Section \ref{sec3_1}) in the RPD. The whole process can be divided into three episodes: the first episode, starting from $R_p \approx  2 R_t$ until the stripped mass reaches a jet-generation threshold, say $\delta M \simeq 0.1 M_\odot $. This episode is driven mainly by the GW emission and lasts long ($\sim$ a few months to years, or $10^{4\sim 6}$ orbits, \citeauthor{2023ApJ...947...32C} \citeyear{2023ApJ...947...32C}). In the second episode, $\delta M>$ 0.1$M_\odot$ and jet is produced, so that the GRB can be observed and there may be QPMs in the GRB light curve. In the third episode, the WD is almost completely destroyed, the debris is dispersed all over along the orbit. The strong accretion and thus the jet are still present, but the QPM is insignificant or disappears.

%%%%%%%%%%%%%%%%%%%%%%%%%%%%%%%%%%%%%%%%%
%%%%%%%%%%%%%%%%%%%%%%%%%%%%%%%%%%%%%%%%%
\section{Searches of QPMs}\label{sec3}

Though it triggered both the GECAM-B \citep{2023GCN.33406....1X} and Fermi-GBM \citep{2023GCN.33407....1D, 2023GCN.33414....1B}, the extremely high brightness of GRB 230307A caused data loss in the time-tagged event data of Fermi-GBM. Therefore, for the light curve analysis of GRB 230307A, we use the GECAM-B data only.

Through visual inspection, we find that the earliest episode (0-14 s) of GRB 230307A's prompt emission appears to show a QPM in its bursting intensity, as shown in Figure \ref{Fig3_1}. With a period of $\sim$ 2.2 s, the modulation lasts for 6 cycles, as marked by the vertical dashed lines in the right panel of Figure \ref{Fig3_1}, and it seems to weaken with time. We also employed several time- and frequency-domain period analysis methods to confirm this QPM (see Appendix \ref{appendix:1}).

For GRBs 060614 and 211211A, we extracted their light curves from the public archives of Swift-BAT and Fermi-GBM, respectively (see Appendix \ref{appendix:1}).

Figure \ref{Fig3_2} shows the light curve of GRB 060614, where in the right panel the enlarged view of the part (-3.5, 54) s since the trigger shows an apparent QPM, with a period of $\sim$ 9.5 s in six cycles. Note that \cite{2006Natur.444.1044G} have first reported a period of 9 s in the $\gamma$-ray light curve of GRB 060614 (time range 7 to 50 s). Subsequently, \cite{2008ApJ...684.1330L} discussed the origin of the $\sim$ 9 s QPM in GRB 060614 in the scenario of a stellar disruption by an intermediate-mass BH. They suggested that the residual material from the disrupted star, moving in a Keplerian orbit, may dominate this QPM.

Similarly, GRB 211211A shows a QPM of period $\approx $ 1.1 s in the early episode of (1, 9) s after the trigger in the $\gamma$-ray light curve, as shown in Figure \ref{Fig3_3}.

%%%%%%%%%%%%%%%%%%%%%%%%%%%%%%%%%%%%%%%%%
%%%%%%%%%%%%%%%%%%%%%%%%%%%%%%%%%%%%%%%%%
\section{Model Application}\label{sec3_1}

Substituting $P$ = 2.2 s into Eq. (\ref{eq4}) and using Eq. (\ref{eq:LebsequeIp1}), we can estimate the mass of the WD to be $M_*$$\sim$ 1.3 $M_\odot$ (here taking $e \simeq 0$). That is, the QPM in the prompt emission episode of GRB 230307A could originate from the RPD of a 1.3 $M_\odot$ WD.

A simple calculation using Eqs. (\ref{eq:LebsequeIp1}-\ref{eq:LebsequeIp2}) and (\ref{eq:LebsequeIp13}) suggest that when the initial stripped mass fraction is $\delta M/M_* \sim 0.1$, the 1.3 $M_\odot$ WD would undergo six consecutive partial disruptions (i.e., 14 s in total) to be completely disrupted, with $\delta M$ at each partial disruption shown in Figure \ref{Fig2_2} (as the second episode there).

\begin{figure}
\centering
\includegraphics[height=4.2cm,width=7.5cm]{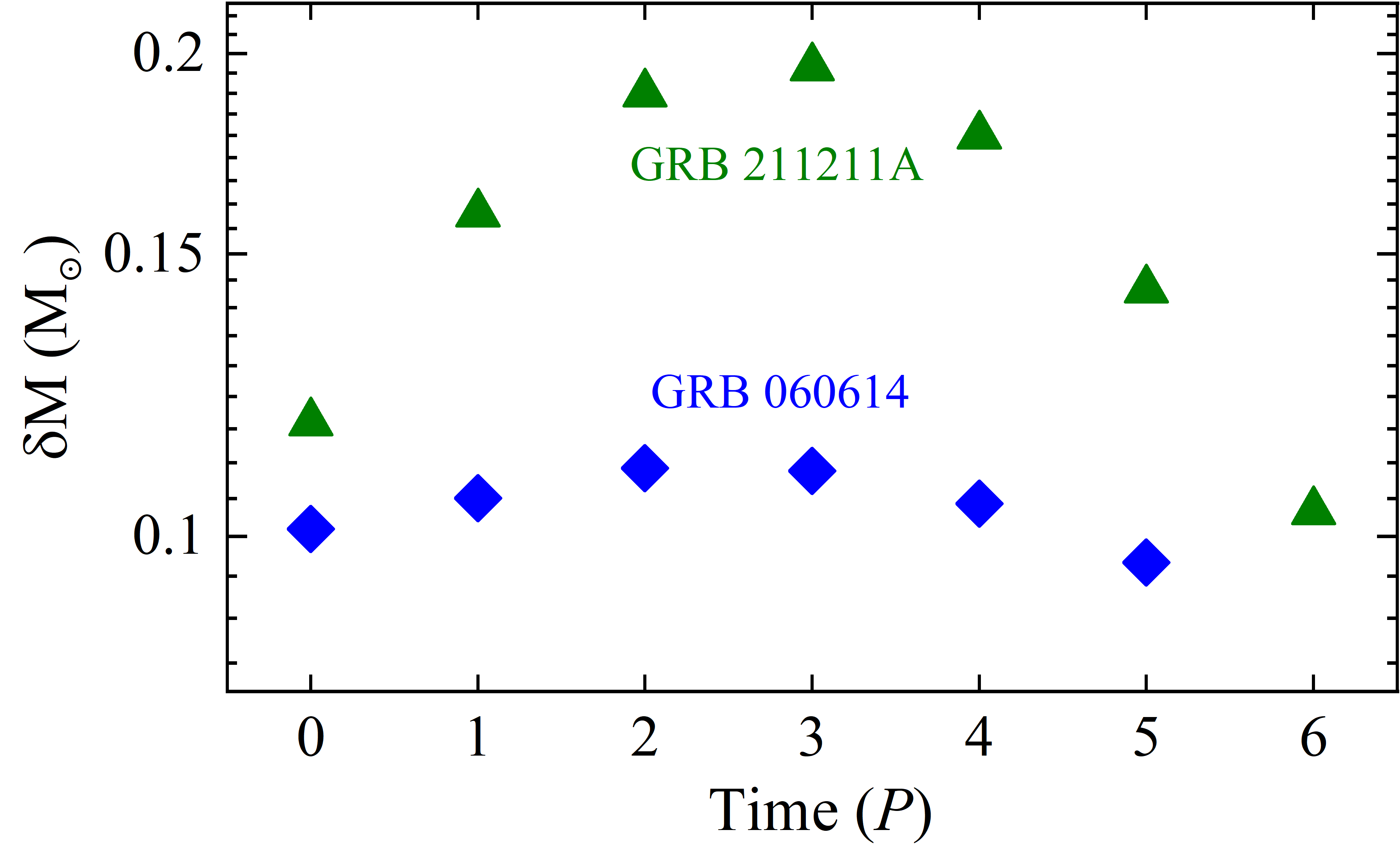}
\caption{The evolution of the stripped masses of the WDs during the QPM episode, in the RPD scenario for GRBs 060614 and 211211A, respectively. The zero time here is chosen to be when the stripped mass firstly reaches 0.1 $M_\odot$.}
\label{Fig4}
\end{figure}

Similarly, the periods of $P$ $\approx$ 9.5 s and $P$ $\approx$ 1.1 s for the QPMs found in GRBs 060614 and 211211A, respectively, would correspond to the WD masses of 0.9 $M_\odot$ and 1.4 $M_\odot$ (close to $M_{ch}$), respectively. Taking $\delta M/M_* \sim 0.1$ as the jet production threshold, the WD is completely destroyed after six and seven cycles of QPM, respectively, consistent with the findings in  Section \ref{sec3}. The variation of $\delta M$ during QPM is displayed in Figure \ref{Fig4}.

The above calculation certainly utilized a few simplifications, notably the WD mass-radius relation (Eq. \ref{eq:LebsequeIp1}) and the way of estimating $\delta M$ (Eq. \ref{eq:LebsequeIp13}). The former was derived from isolated WDs and may not be applicable to the WD in dynamical situations like in RPDs. However, the general trend of increasing $R_*$ with decreasing $M_*$ would probably remain, while tidal heating may also contribute to this. The latter was actually derived for shallow strippings ($\delta M/M_* \ll 1$); whether it could work for cases of $\delta M \rightarrow M_*$ is questionable. In addition, how to translate from the $\delta M$ evolution to the light curve is complicated by how the central object digests the stripped mass and how much fraction of the power goes to the jet, etc. All these deserve future investigation.

The QPM parameters for GRBs 230307A, 060614, and 211211A are summarized in Table \ref{tab1}, and all data analysis procedures are included in Appendix \ref{appendix:1}.

\begin{table} 
    \centering 
    \caption{The QPMs parameters for GRBs 230307A, 060614 and 211211A}
    \label{tab1} 
    \renewcommand\arraystretch{1.2}  
    \setlength{\tabcolsep}{2mm} 
    \begin{tabular}{lccc} %four columns, alignment for eac 
\hline 
Properties         &  230307A&  060614& 211211A\\ 
\hline
%Redshift           &  0.065         &  0.125           &  0.076          \\
%Offset (kpc)       &  36.6          &  0.80            &  8.19          \\
$T_{90}$ (s)       &  42            &  102             &  43             \\
$P$ (s)              &  2.2 $\pm$ 0.16&   9.5 $\pm$ 0.6  &   1.1 $\pm$ 0.1\\
Time span (s)      &  0 - 14&  -3.5 - 54&  1 - 9\\
number of cycles   &  $\sim$ 6      &  $\sim$ 6        &  $\sim$ 7       \\
$M_*$ ($M_\odot$)  & $\sim$ 1.3     & $\sim$ 0.9       & $\sim$ 1.4      \\
\hline 
    \end{tabular} 
\end{table}

%%%%%%%%%%%%%%%%%%%%%%%%%%%%%%%%%%%%%%%%%
%%%%%%%%%%%%%%%%%%%%%%%%%%%%%%%%%%%%%%%%%
\section{Conclusion and discussion}
\label{sec4}
This work studies the process of RPD of WDs in WD-NS or WD-BH mergers. Due to the larger $R_t$ of the WD, and thus a probable residual orbital eccentricity, RPDs are almost inevitable. If WD-NS or WD-BH mergers can generate GRBs, we predict that a QPM may be detected in the light curve of the GRB.

The progenitors of a special subgroup of GRBs, such as GRBs 060614, 211211A, and 230307A, remain as a puzzle. These GRBs have long durations, large offsets from the host galaxies and the follow-up observations have no supernovae associated with them but are related to potential kilonovae. It is natural to associate their progenitors with WD-NS or WD-BH mergers.

We extracted the light curves of GRBs 230307A, 060614, and 211211A from the observational data of GECAM-B, Swift-BAT, and Fermi-GBM, respectively. We find visible QPMs in all three GRBs, with the periods of 2.2 s, 9.5 s and 1.1 s, corresponding to the WD mass of 1.3 $M_\odot$, 0.9 $M_\odot$ and 1.4 $M_\odot$, respectively. 

GRBs 230307A and 211211A have very short QPM periods ($\sim$ 2.2 s and $\sim$ 1.1 s, respectively), which require extremely dense (thus massive, according to Eq. \ref{eq:LebsequeIp1}) WDs so that $R_t$ is small and the orbital period is short. A few near-Chandrasekhar-mass-limit ($M_*\sim 1.3-1.4 M_{\odot}$) WDs have been observationally verified (e.g., \citeauthor{1995MNRAS.277..971B} \citeyear{1995MNRAS.277..971B}, \citeauthor{2010A&A...524A..36K} \citeyear{2010A&A...524A..36K}, \citeauthor{2020MNRAS.499L..21P} \citeyear{2020MNRAS.499L..21P}, \citeauthor{2021Natur.595...39C} \citeyear{2021Natur.595...39C}), and how they were formed and remained stable is a puzzle. They are found to be hot, rapidly rotating and highly magnetized, which may points to binary-WD merger as the likely formation channel and magnetism may play a role in stabilizing them (e.g., \citeauthor{2014MNRAS.438...14D} \citeyear{2014MNRAS.438...14D})

The statistical significances of the QPMs found in the three GRBs are not very high (see Appendix \ref{appendix:1} ). This may be attributed to the enormous complexity involved in the actual process, including the mass transfer and accretion onto the compact companion, as well as jet launching and energy dissipation within the jet and radiating. However, the presence of QPMs aligns with our model predictions, providing supports for WD-NS or WD-BH as the progenitor(s) of this sub-type of GRBs.

Note that one condition for the modulation of GRB flux through the RPD of a WD is the presence of a remaining $e$. However, in an extreme case where the initial $e$ was already low when the compact object binary was formed, the remaining $e$ would be very close to 0 upon the merger. It would still be possible to generate an accretion disk, resulting in a relativistic jet and a GRB. However, in such cases, the significance of QPM would be low or absent. Another piece of uncertainty may be related to the sucessfulness of launching a GRB jet, due to the relatively low magnetization of the stripped WD material. This may explain the currently small sample size of this GRB subgroup.

In our model, after the complete disruption of the WD, the fast accretion of its remains not only generates a jet but also launches a slower outflow. The collision between the outflowing material and the surrounding debris may manifest as the kilonova emission. Alternatively, the heating from the radioactive decay of $^{56}$Ni within the outflow \citep{2012MNRAS.419..827M, 2022MNRAS.510.3758B, 2023ApJ...947L..21Z, 2024ApJ...964L...9W} is the source of radiation for the so-called kilonova component.

%%%%%%%%%%%%%%%%%%%%%%%%%%%%%%%%%%%%%%%%%
%%%%%%%%%%%%%%%%%%%%%%%%%%%%%%%%%%%%%%%%%
\section*{Acknowledgments}
We thank Long Ji, Bing Zhang, Bing Li and Yacheng Kang for their helpful discussions, and an anonymous reviewer for providing constructive feedback. This work is supported by the National Natural Science Foundation of China (grants 12073091, 12261141691 and 12393814). We acknowledge the data resources from “National Space Science Data Center, National Science \& Technology Infrastructure of China (\url{https://www.nssdc.ac.cn})”. 

%%%%%%%%%%%%%%%%%%%%%%%%%%%%%%%%%%%%%%%%%
%%%%%%%%%%%%%%%%%%%%%%%%%%%%%%%%%%%%%%%%%
\bibliography{RPD_GRBs}{}
\bibliographystyle{aasjournal}

\appendix
\section{Methods}
\label{appendix:1}
\renewcommand{\thefigure}{A\arabic{figure}}
\subsection{Observation data}

We extracted GRB 230307A's observational data from the GECAM-B archive and generated multiple light curves in different energy bands (30-100 keV, 100-300 keV, 300-500 keV, 500 keV-2000 keV, and 30-2000 keV) at a time resolution of 64 ms.

For GRB 060614, we obtained the observational data from Swift-BAT's public archive \url{https://www.swift.ac.uk/archive/ql.php}. Using HEASoft tools (version HEASoft 6.30) and following standard analysis procedures (\url{https://www.swift.ac.uk/analysis/bat/index.php}), we extracted two light curves covering the energy range of 15-350 keV, with time resolutions of 64 ms and 256 ms.

We retrieved the time-tagged event (TTE) data set for GRB 211211A from Fermi-GBM's public data archive (\url{https://heasarc.gsfc.nasa.gov/W3Browse/fermi/fermigbrst.html}). Two sodium iodide (NaI) detectors, n2 and na, closest to the source direction were selected. By utilizing the Fermi GBM Data Tools (\url{https://fermi.gsfc.nasa.gov/ssc/data/analysis/gbm}, version Fermi GBM Data Tools: v1.1.1), we extracted a light curve covering the energy range of 15-5000 keV, with a time resolution of 64 ms.

\subsection{QPM identification}
To identify and confirm the QPMs in the light curve, we employed commonly used astronomical time-domain analysis methods, including the Weighted Wavelet Z-Transform (WWZ), Power Density Spectrum (PDS), and Lomb-Scargle Periodogram (LSP). Now we specifically the process of QPM identification in the light curve of GRB 230307A in the energy range of 500-2000 keV.

The WWZ method is widely used in the field of periodic search, which searches for possible periodic in a time series by varying the parameters of the wavelet to fit a light curve in the time-frequency domain \citep{1996AJ....112.1709F}. Unlike the traditional Fourier variation, WWZ can capture the frequency and the temporal location at different temporal resolutions so that it can identify transient periodic lurking at specific locations in the time series data.

\begin{figure}
\centering
\includegraphics[height=5.6cm,width=7.4cm]{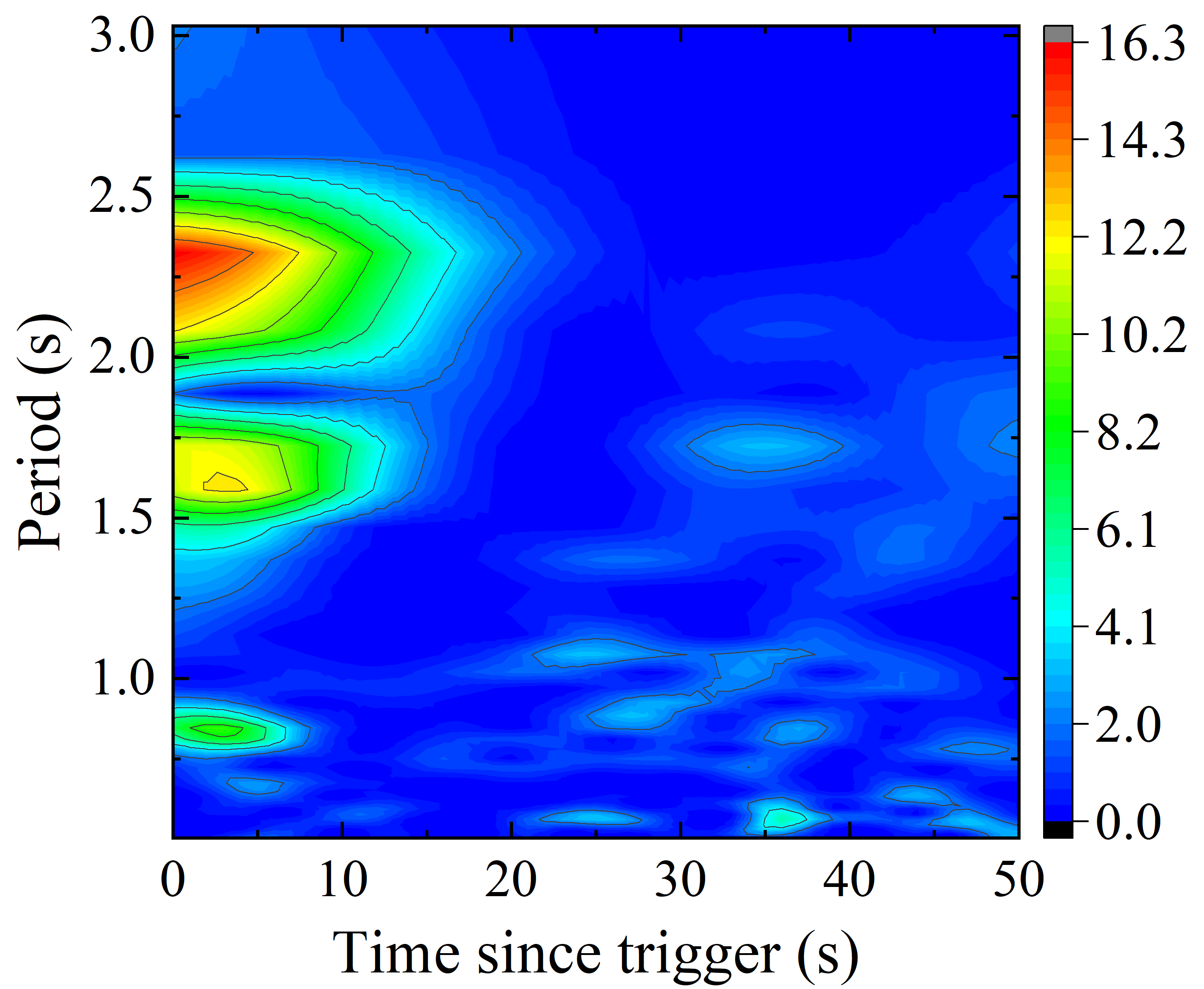}
\caption{The Weighted Wavelet Z-transform (WWZ) power plot of the light curve (0-50 s) of GRB 230307A in the 500-2000 keV band.}
\label{Fig9}
\end{figure}

Therefore, we initially employed the WWZ method to determine the occurrence of QPM in the light curve. Figure \ref{Fig9} demonstrates the evolution of WWZ power with time and frequency. From figure, we can observe that within the 0-14 s range, there is a significant WWZ power peak between 2 s and 2.5 s, indicating the presence of potential QPMs.

\begin{figure*}
\begin{minipage}[t]{1\textwidth}
\centering
\includegraphics[height=5.25cm,width=8.7cm]{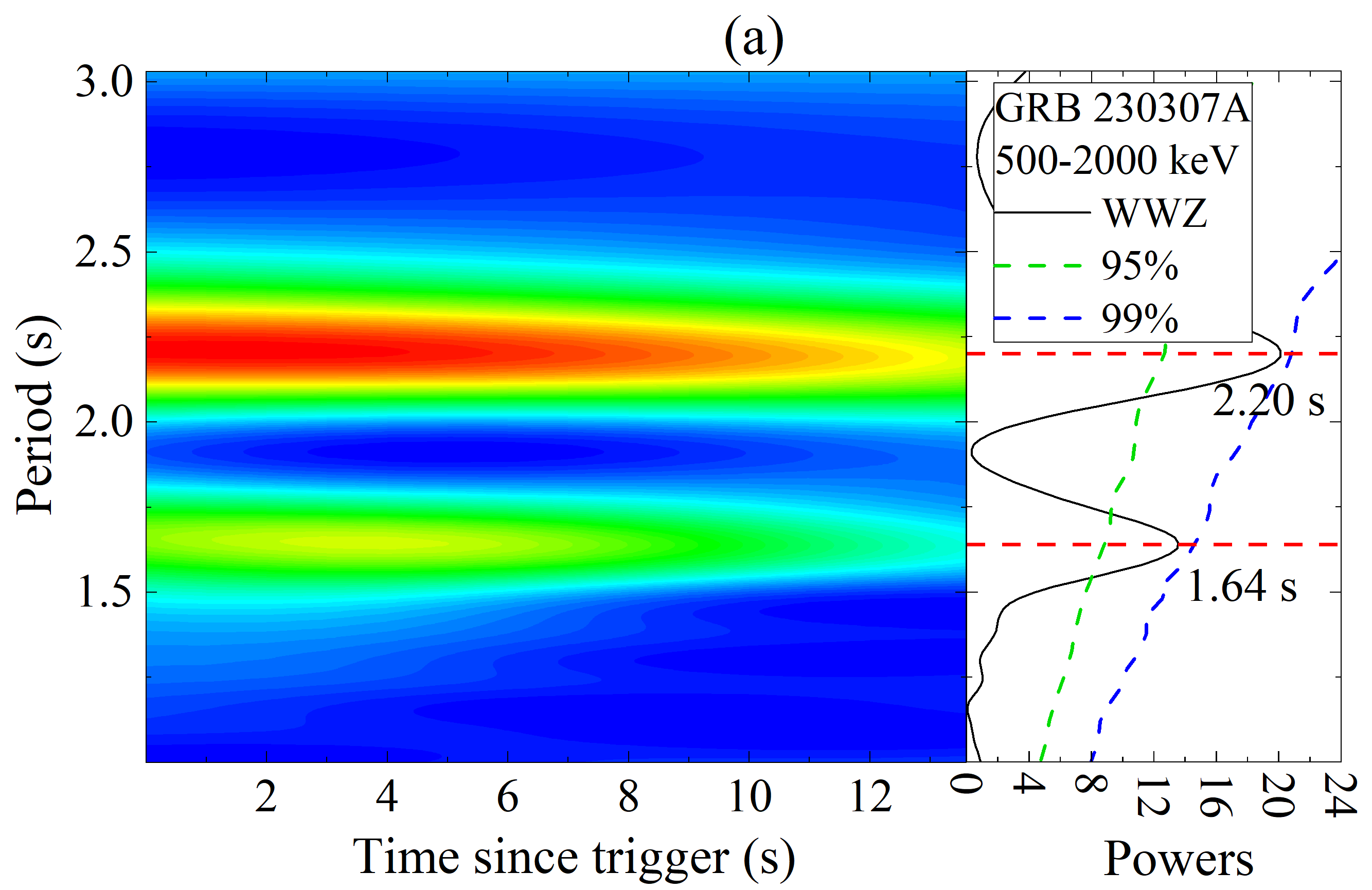}
\includegraphics[height=5.2cm,width=5.9cm]{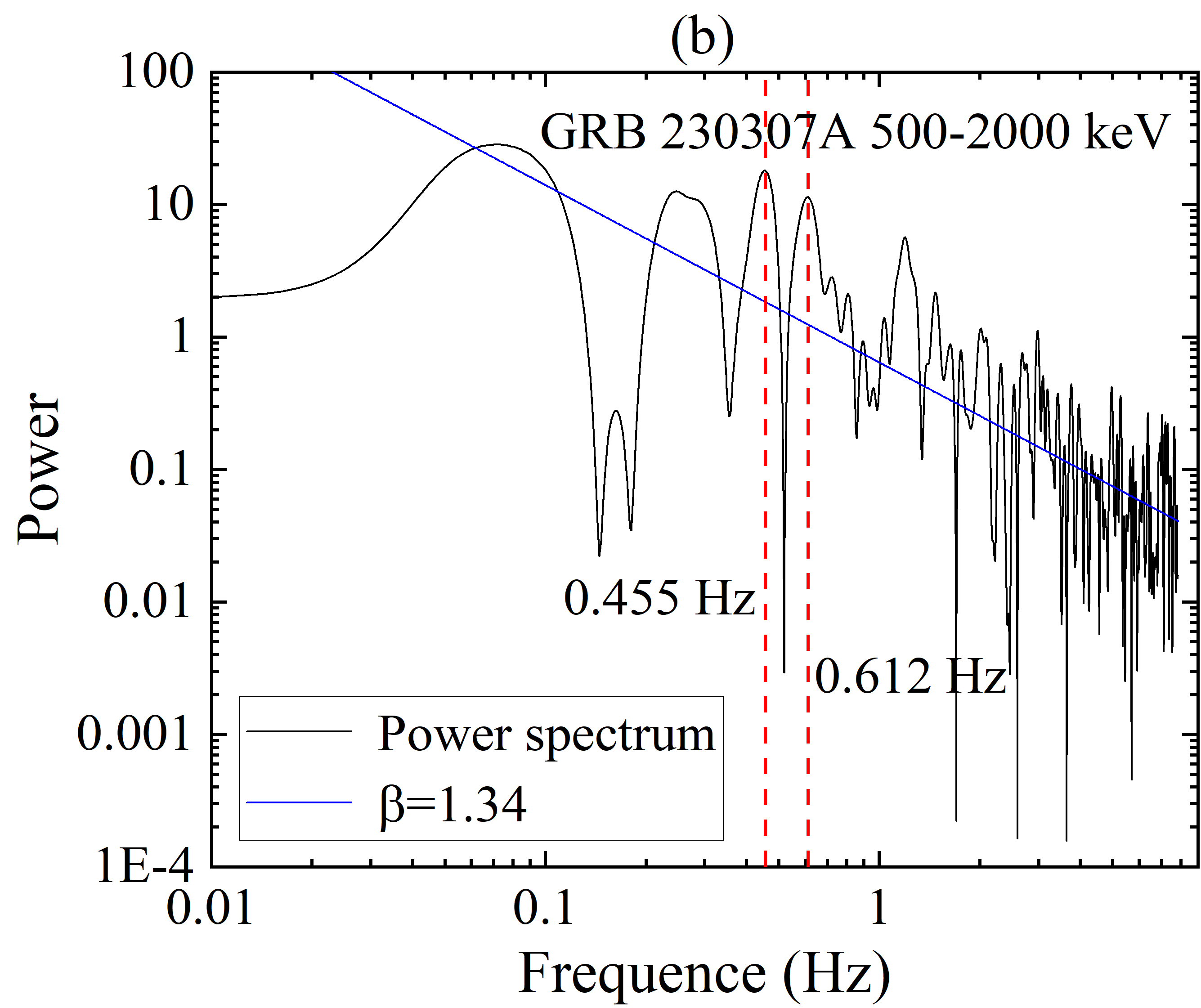}
\centering
\includegraphics[height=5.1cm,width=6.5cm]{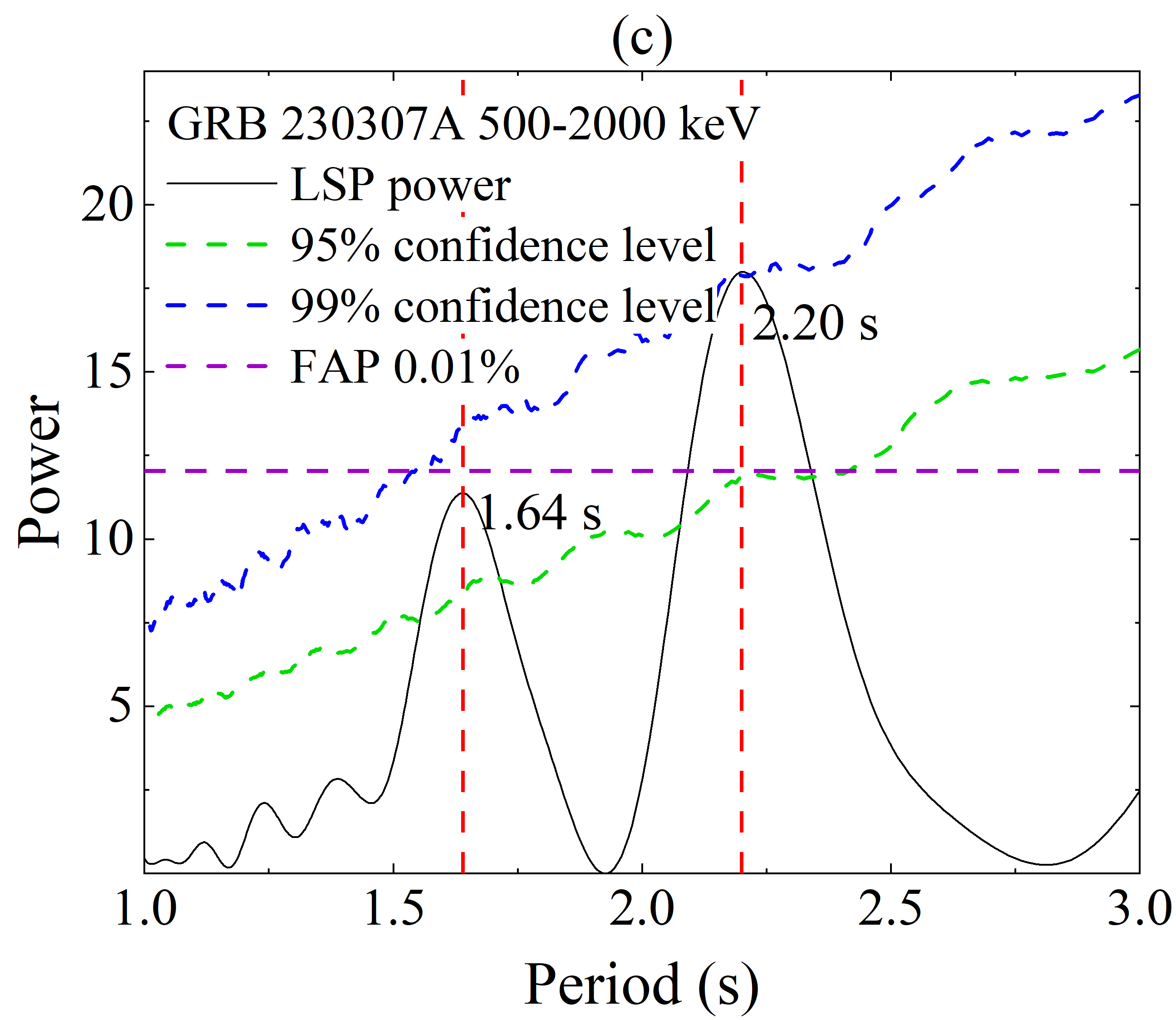}
\includegraphics[height=5.1cm,width=6.5cm]{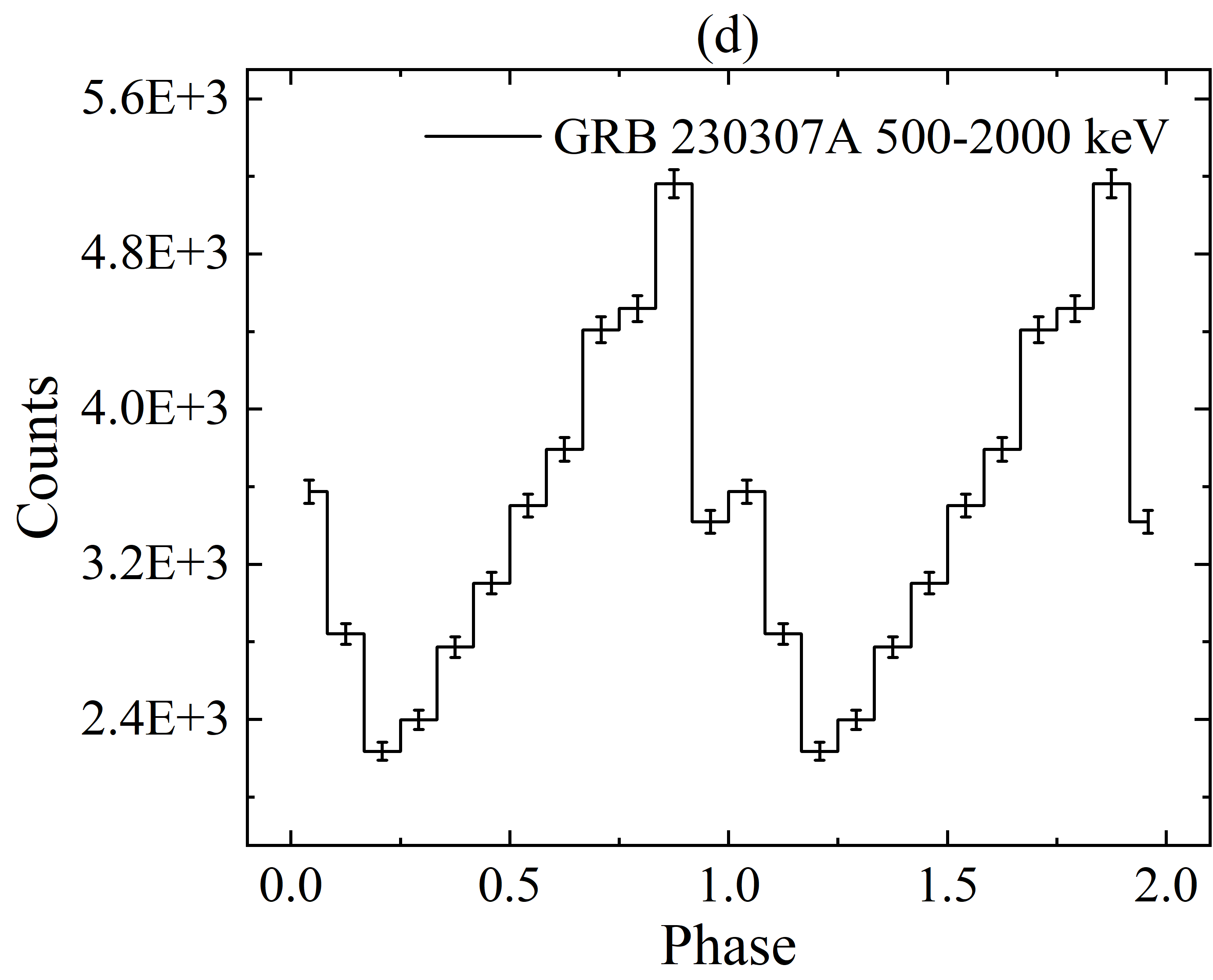}
\end{minipage}
\caption{The periodic analysis results of the light curve of GRB 230307A in the QPM phase in the 500-2000 keV band. Panels a, b, c and d show the results of WWZ, PDS, LSP and epoch folding, respectively. The red dashed line indicates the period corresponding to the extreme peak power, and the green, blue, and purple dashed lines represent the 95\%, 99\% and $FAP$=0.01\% confidence levels, respectively.}
\label{fig10}
\end{figure*}

Next, we will focus on the QPM episode (0-14 s). Panel (a) of Figure \ref{fig10} displays the WWZ power of the light curve during the QPM episode. The solid black line represents the time-averaged WWZ power, while the dashed red line represents the period (2.20$\pm$0.16 s) corresponding to the peak power. Note that the uncertainty in all results was estimated as the half-height and half-width of the Gaussian function of the fitted power peak.

To minimize false detections from a single method, we performed a Fourier transform on the QPM phase and obtained its PDS, as shown in panel (b) of Figure \ref{fig10}. The PDS exhibits a power-law distribution (with a fitted $\beta$ = 1.34) and a peak frequency of $\sim$ 0.455. Additionally, panel (c) of Figure \ref{fig10} presents the results of the LSP analysis, which is an improved method based on the Fourier transform (for more detailed information, see \citeauthor{1976Ap&SS..39..447L} \citeyear{1976Ap&SS..39..447L}, \citeauthor{1982ApJ...263..835S} \citeyear{1982ApJ...263..835S}, \citeauthor{2018ApJS..236...16V} \citeyear{2018ApJS..236...16V}). The peak power in the LSP corresponds to a period of 2.20$\pm$0.18. These results are consistent with the findings from WWZ and PDS analyses.

\begin{table*}
   \centering
   \caption{The results obtained from the analysis of QPMs in GRB using different methods are as follows: (Note that the confidence level for MC simulations is taken here to be the highest value given by several methods.)}
   \label{tabA1}
   \renewcommand\arraystretch{1.2}
  \setlength{\tabcolsep}{2mm}
   \begin{tabular}{lccccc} % four columns, alignment for each
\hline
GRB                      &  WWZ (s)         & PDS (Hz)    &   LSP (s)   &   FAP  &   MC confidence level  \\
\hline
230307A  (500-2000 keV) &  2.20$\pm$0.16 & 0.455 ($\beta=1.34$)& 2.20$\pm$0.18 &  $<$ 0.01\% &  $\sim$ 99\%   \\
                        &  1.64$\pm$0.12 & 0.612               & 1.64$\pm$0.11 &  $>$ 0.01\% &  $\sim$ 98\%   \\
230307A (300-500 keV)   &  2.22$\pm$0.15 & 0.452 ($\beta=1.36$)& 2.21$\pm$0.17 &  $<$ 0.01\% &  $\sim$ 98\%   \\
                        &  1.63$\pm$0.10 & 0.611               & 1.63$\pm$0.11 &  $>$ 0.01\% &  $\sim$ 97.5\% \\
230307A (100-300 keV)   &  2.23$\pm$0.12 & 0.447 ($\beta=1.3$) & 2.23$\pm$0.14 &  $<$ 0.01\% &  $\sim$ 97\%   \\
                        &  1.64$\pm$0.11 & 0.605               & 1.65$\pm$0.12 &  $>$ 0.01\% &  $\sim$ 96.5\% \\
060614 (15-350 keV)     &  9.53$\pm$0.63 & 0.105 ($\beta=1.3$) & 9.52$\pm$0.6  &  $<$ 0.01\% &  $\sim$ 99.8\% \\
211211A (15-5000 keV)   &  1.10$\pm$0.16 & 0.92 ($\beta=1.58$) & 1.08$\pm$0.06 &  $<$ 0.01\% &  $\sim$ 97\%   \\
\hline
    \end{tabular}
\end{table*}

\subsection{Significance estimate}
Any significant periodic variations are dependent on assumptions made about pseudo-random processes that mimic the periodic variations \citep{2015MNRAS.449.3293L}. In the case of GRB 230307A, the limited number of cycles in the transient QPM makes it difficult to estimate the significance of the QPM. 

We assessed the significance of the power peaks using the false alarm probability (FAP), $FAP(P_n)=1- (1-prob(P>P_n))^M$, where $FAP$ denotes the likelihood that at least one of the independent power values at a given frequency in the white noise periodogram is greater than or equal to the power threshold $P_n$ \citep{2018ApJS..236...16V}. The purple dashed line in panel (c) of Figure \ref{fig10} shows that with $FAP$ = 0.01\%, the significance at peak power (P = 2.2 s) is higher than 4$\sigma$. Indeed, it has been established that the Power Density Spectrum (PDS) of the light curve may follow a power-law distribution (As shown in panel (b) of Figure \ref{fig10}), resembling the spectral shape of red noise. We address this issue using two approaches.

(1) Assuming that the red noise is generated by flares of similar amplitudes, we studied the coherence of periodic modulations by examining the phase variations of the light curve. We folded the light curve phase with a period of 2.2 s to minimize the influence of red noise on the process of period modulation. The errors were calculated with a 1-sigma uncertainty, and the zero phase corresponds to the start time of the QPM episode. We set 12 phase bins to display the phase variations. The results are shown in panel (b) of Figure \ref{fig10}. From figure, it can be observed that within 2.2 s, the rising and falling trends of the light curve are asymmetric, and there is a significant phase modulation in the flux variations.

(2) To evaluate the significance of the QPM, we generated synthetic light curves through Monte Carlo (MC) simulations, with artificial light curves having the same PSD and probability density function as the original light curve \citep{1995A&A...300..707T, 2013MNRAS.433..907E}. We simulated $10^4$ synthetic light curves for the GRB light curve. The green and blue dashed lines in panels (a) and (c) of Figure \ref{fig10} represent the 95\% and 99\% confidence contour levels, respectively. The QPM significance at $\sim$ 2.2 s is approximately 99\%, indicating a probability of less than 1\% for the presence of statistical fluctuations.

In addition to the 2.2-s period at the power peak, we also observed a periodicity of 1.6 s, although its power is lower compared to the former. Interestingly, we found that the ratio between these two periods is similar to the 3:2 period ratio observed in X-ray binary systems.

\begin{figure*}
\begin{minipage}[t]{1\textwidth}
\centering
\includegraphics[height=5.2cm,width=8.7cm]{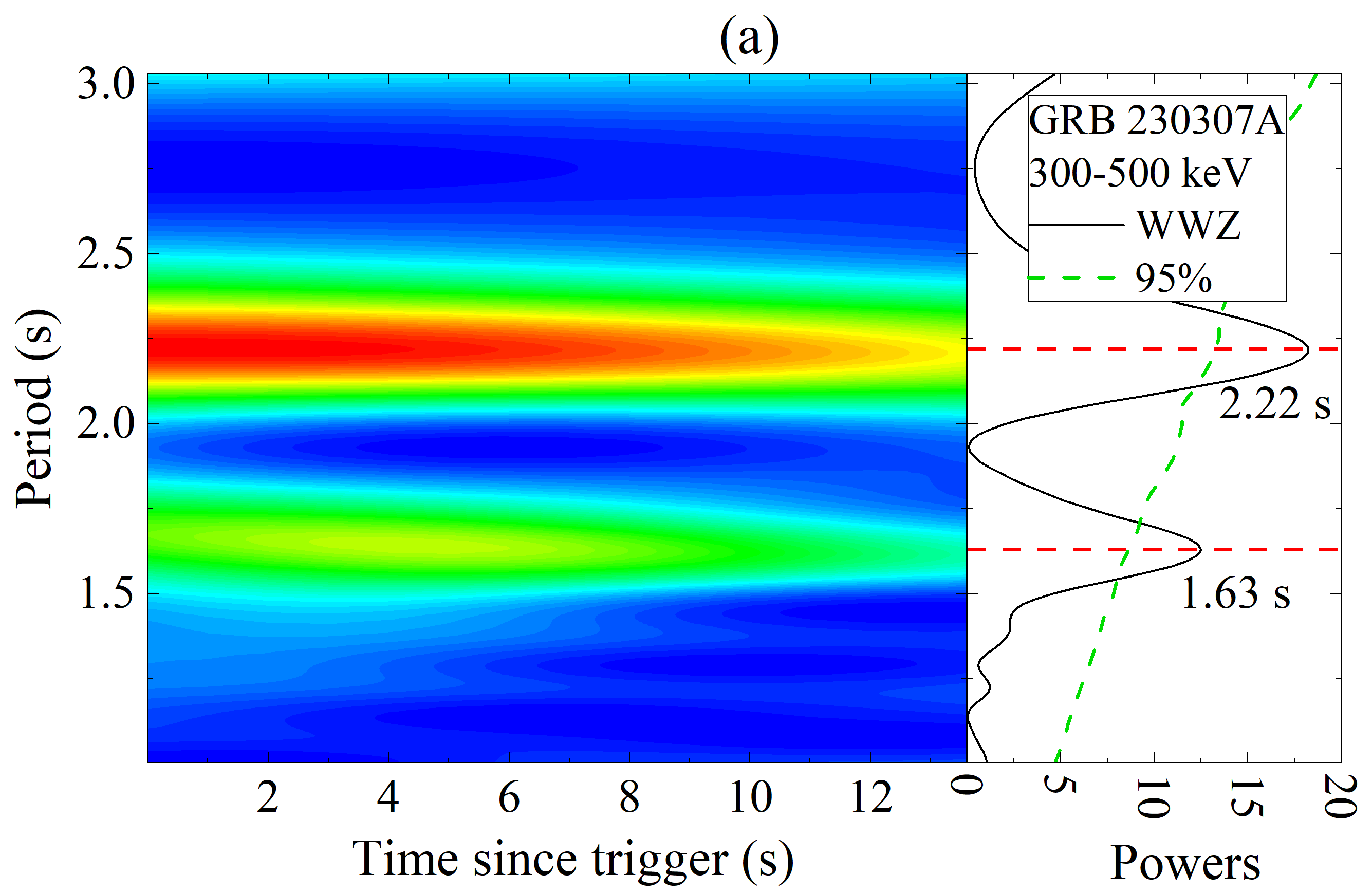}
\includegraphics[height=5.2cm,width=5.9cm]{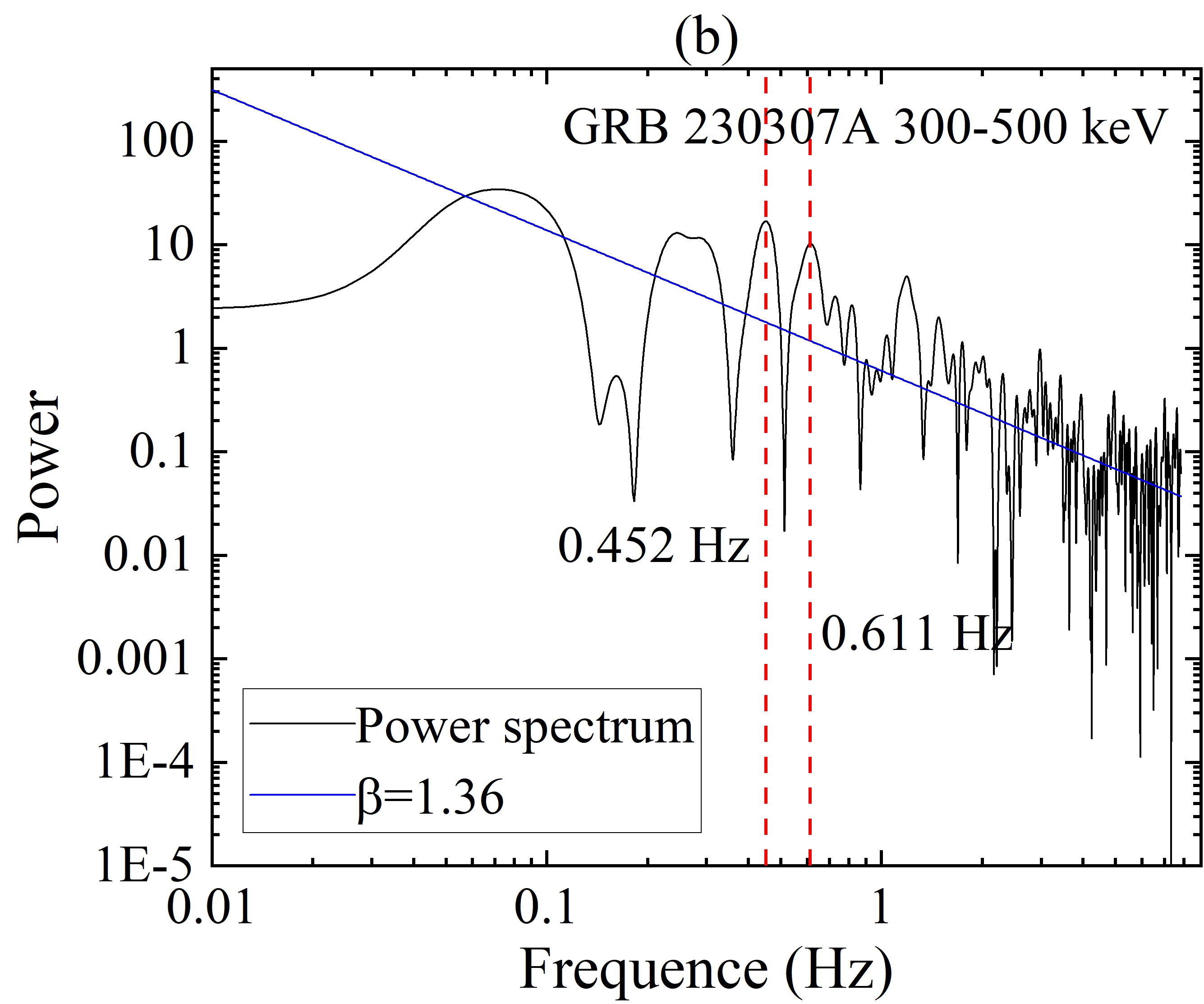}
\centering
\includegraphics[height=5.1cm,width=6.5cm]{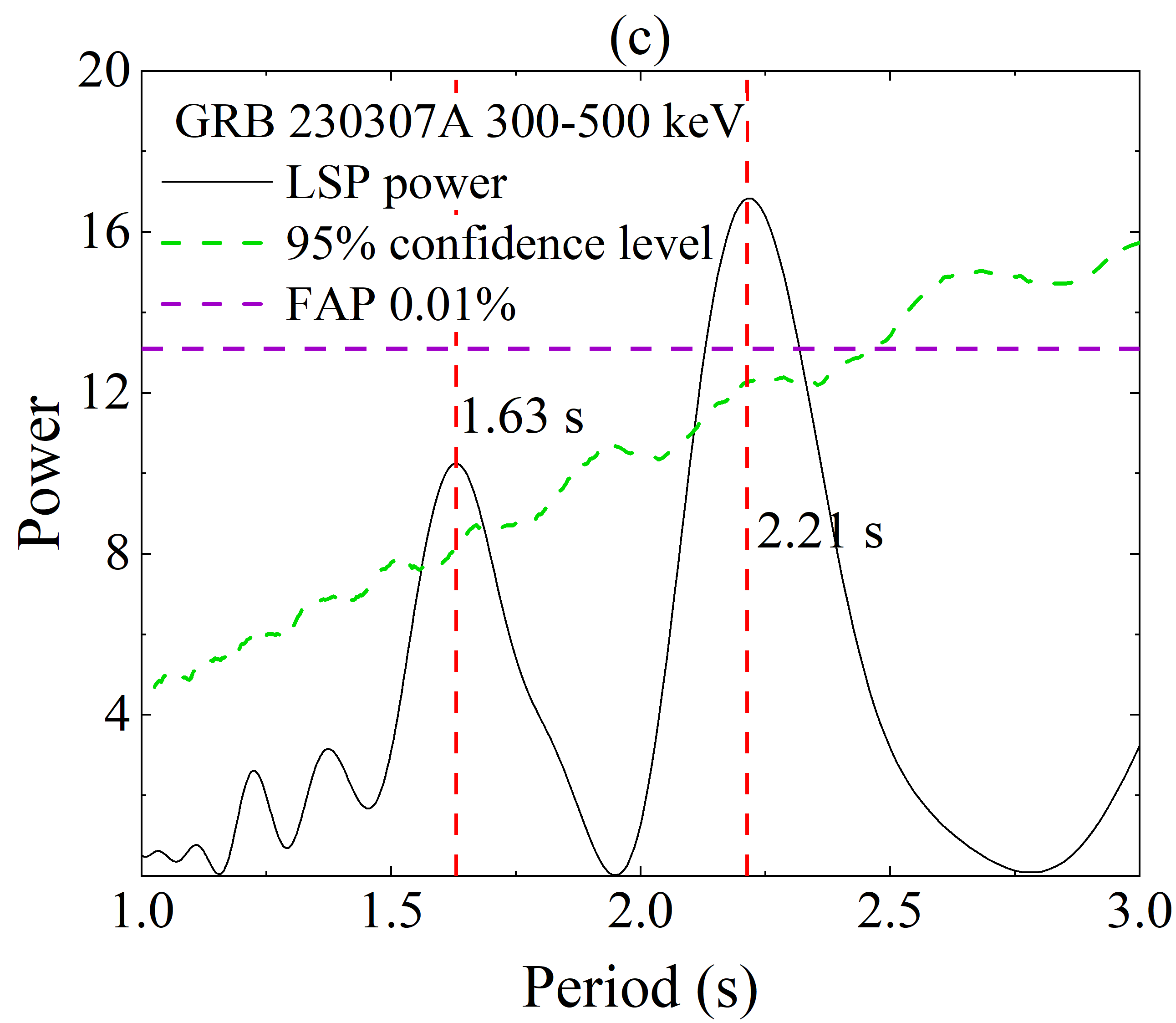}
\includegraphics[height=5.1cm,width=6.5cm]{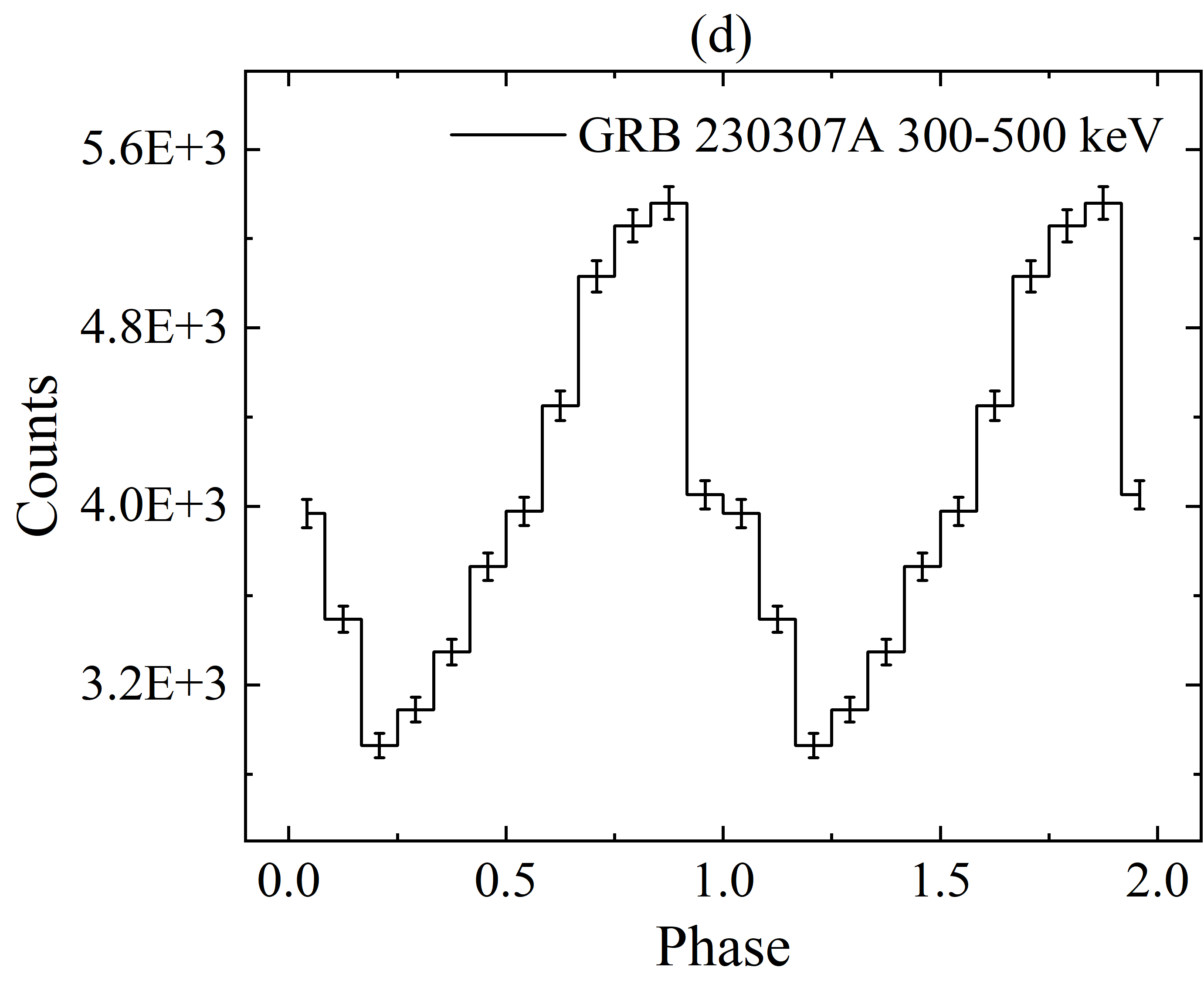}
\end{minipage}
\caption{Same as in Figure \ref{fig10} but for GRB 230307A (300-500 keV).}
\label{fig11}
\end{figure*}

\begin{figure*}
\begin{minipage}[t]{1\textwidth}
\centering
\includegraphics[height=5.2cm,width=8.7cm]{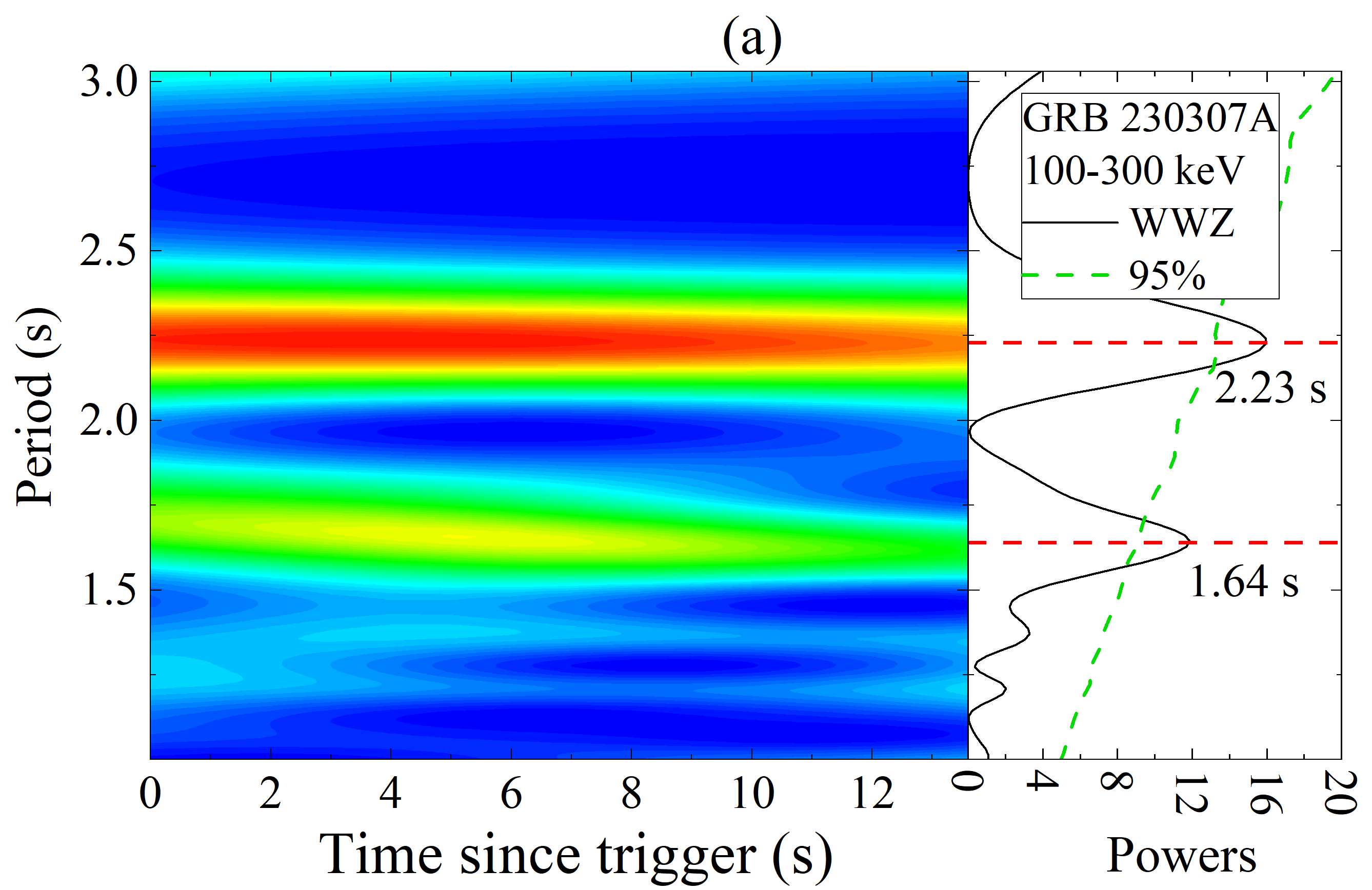}
\includegraphics[height=5.2cm,width=5.9cm]{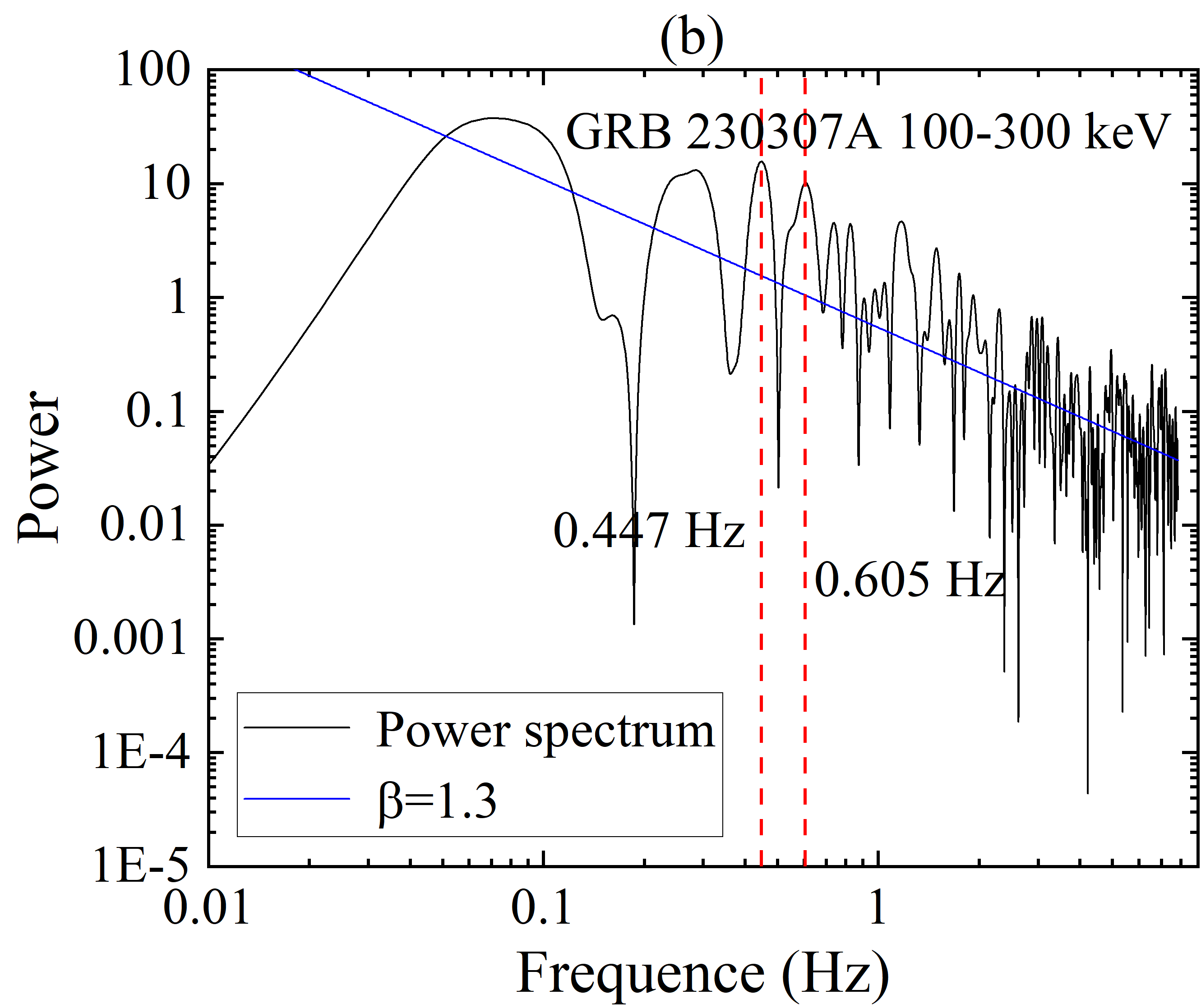}
\centering
\includegraphics[height=5.1cm,width=6.5cm]{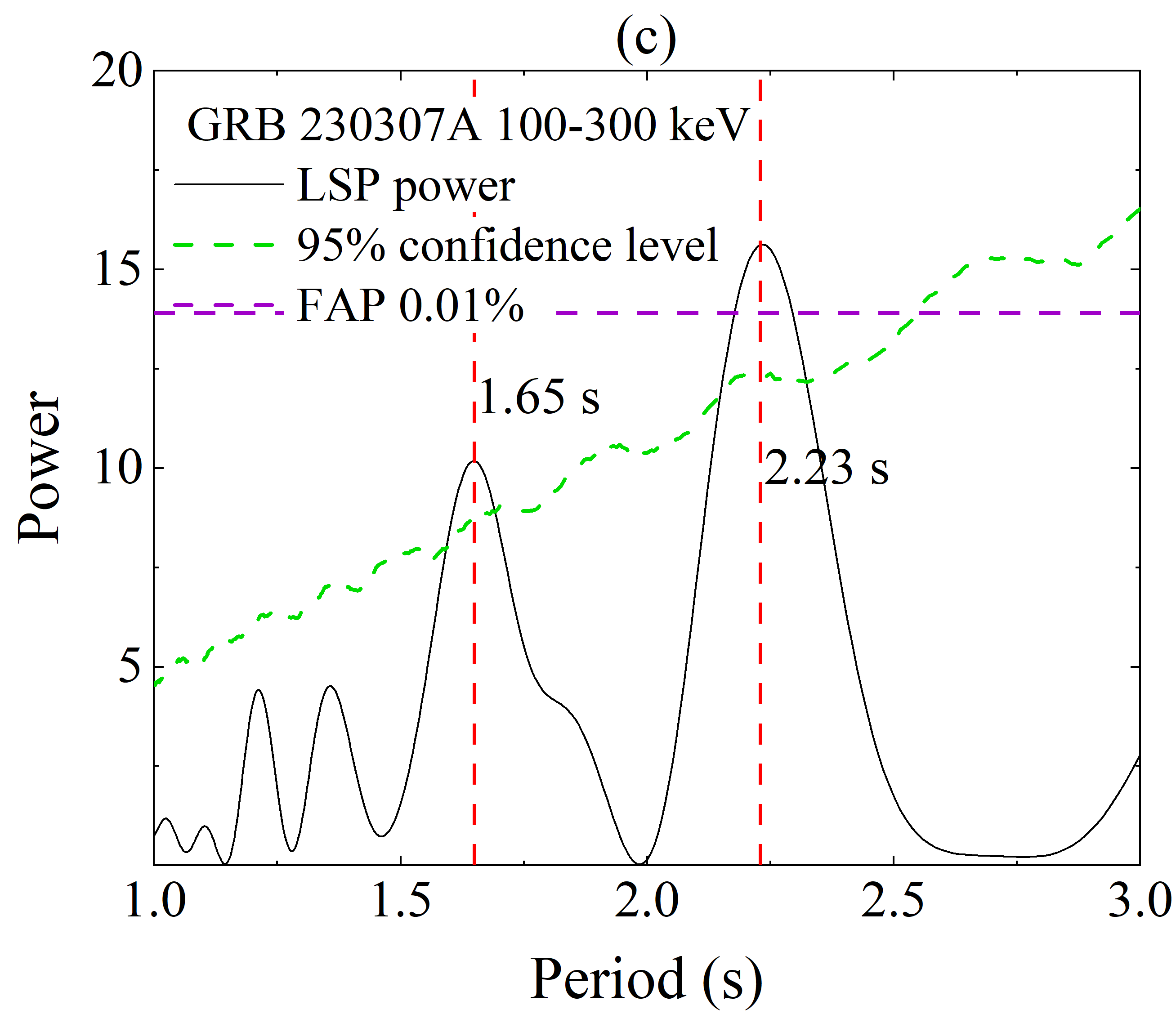}
\includegraphics[height=5.1cm,width=6.5cm]{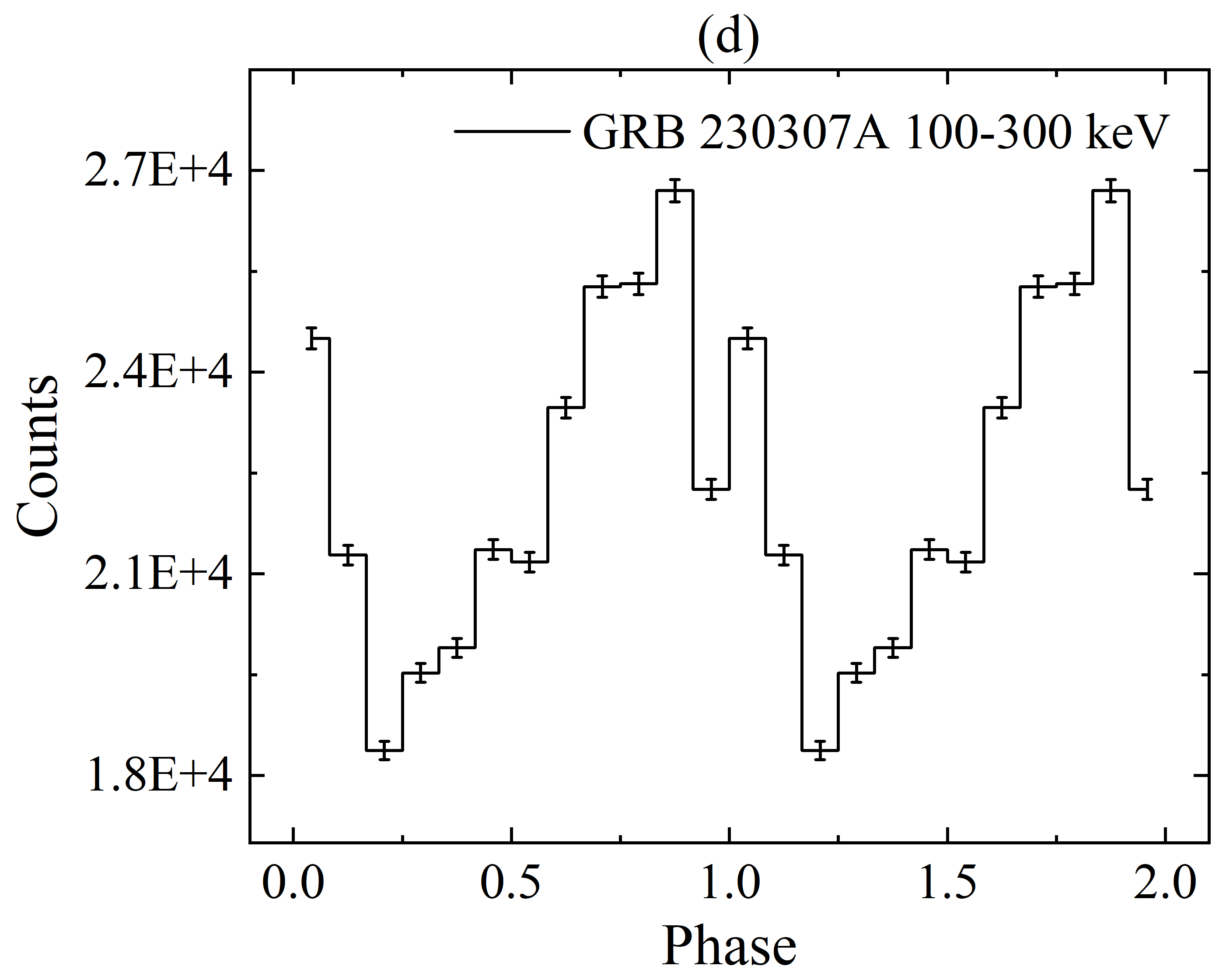}
\end{minipage}
\caption{Same as in Figure \ref{fig10} but for GRB 230307A (100-300 keV).}
\label{fig12}
\end{figure*}

\begin{figure*}
\begin{minipage}[t]{1\textwidth}
\centering
\includegraphics[height=5.2cm,width=8.7cm]{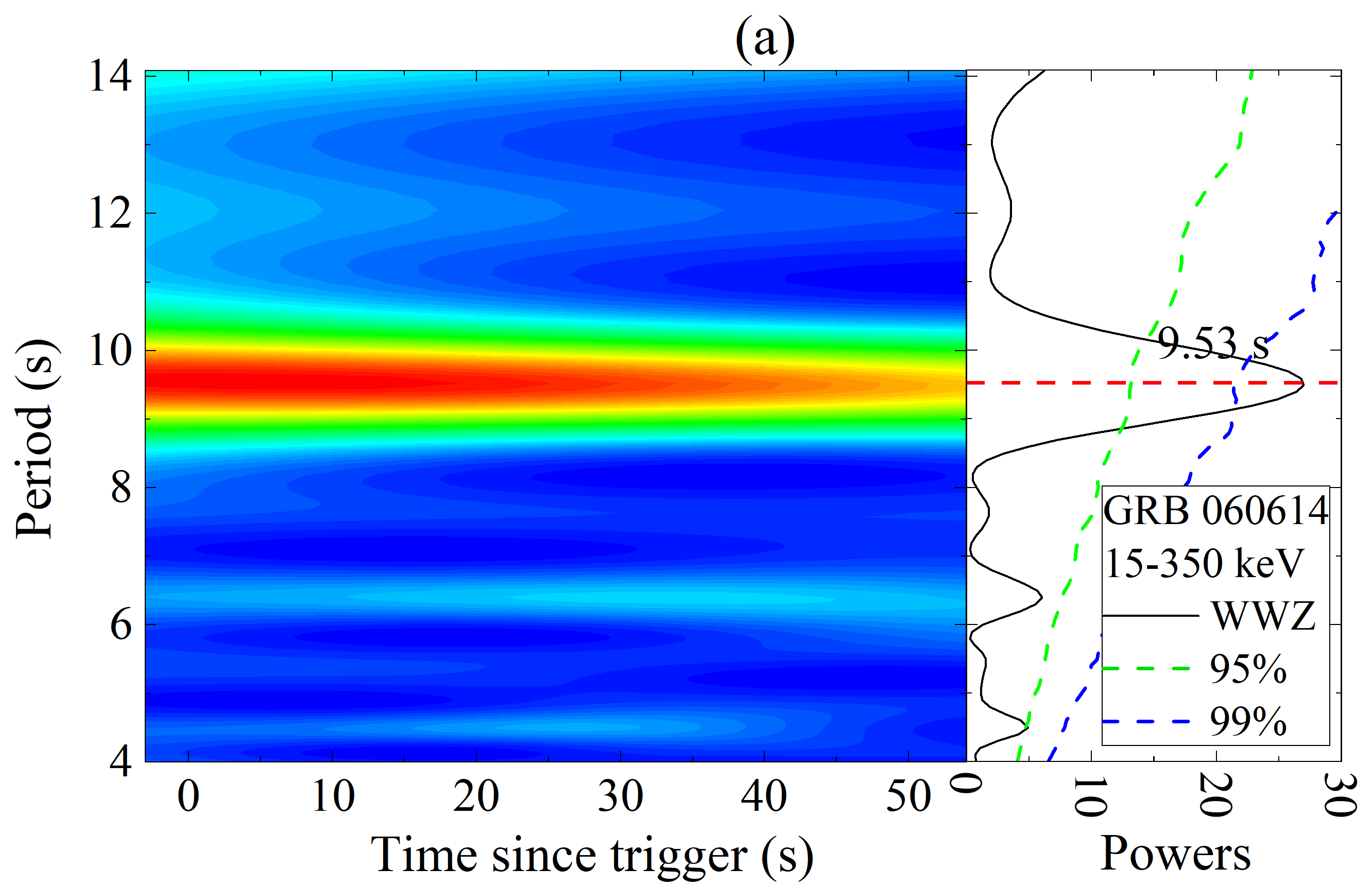}
\includegraphics[height=5.2cm,width=5.9cm]{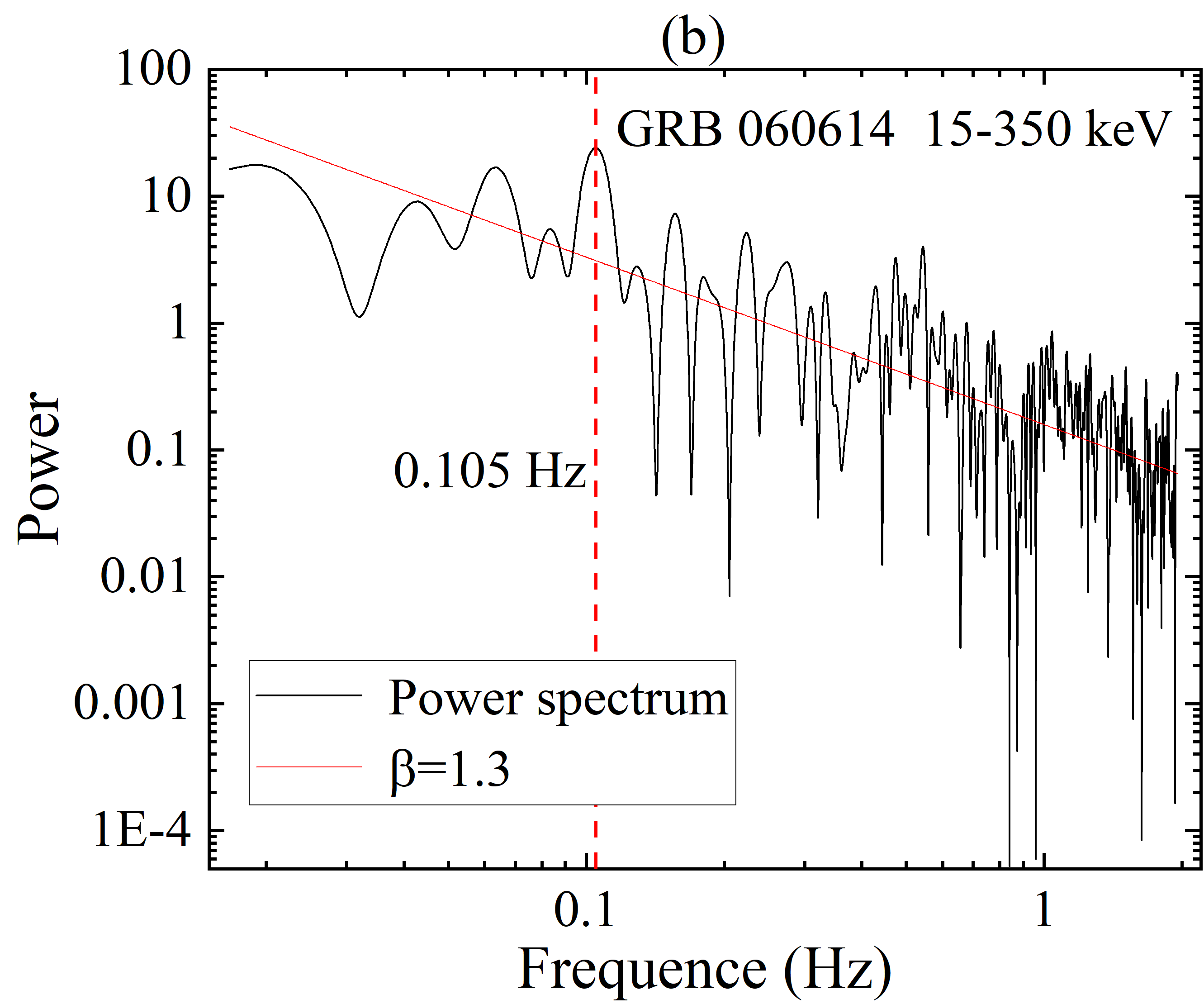}
\centering
\includegraphics[height=5.1cm,width=6.5cm]{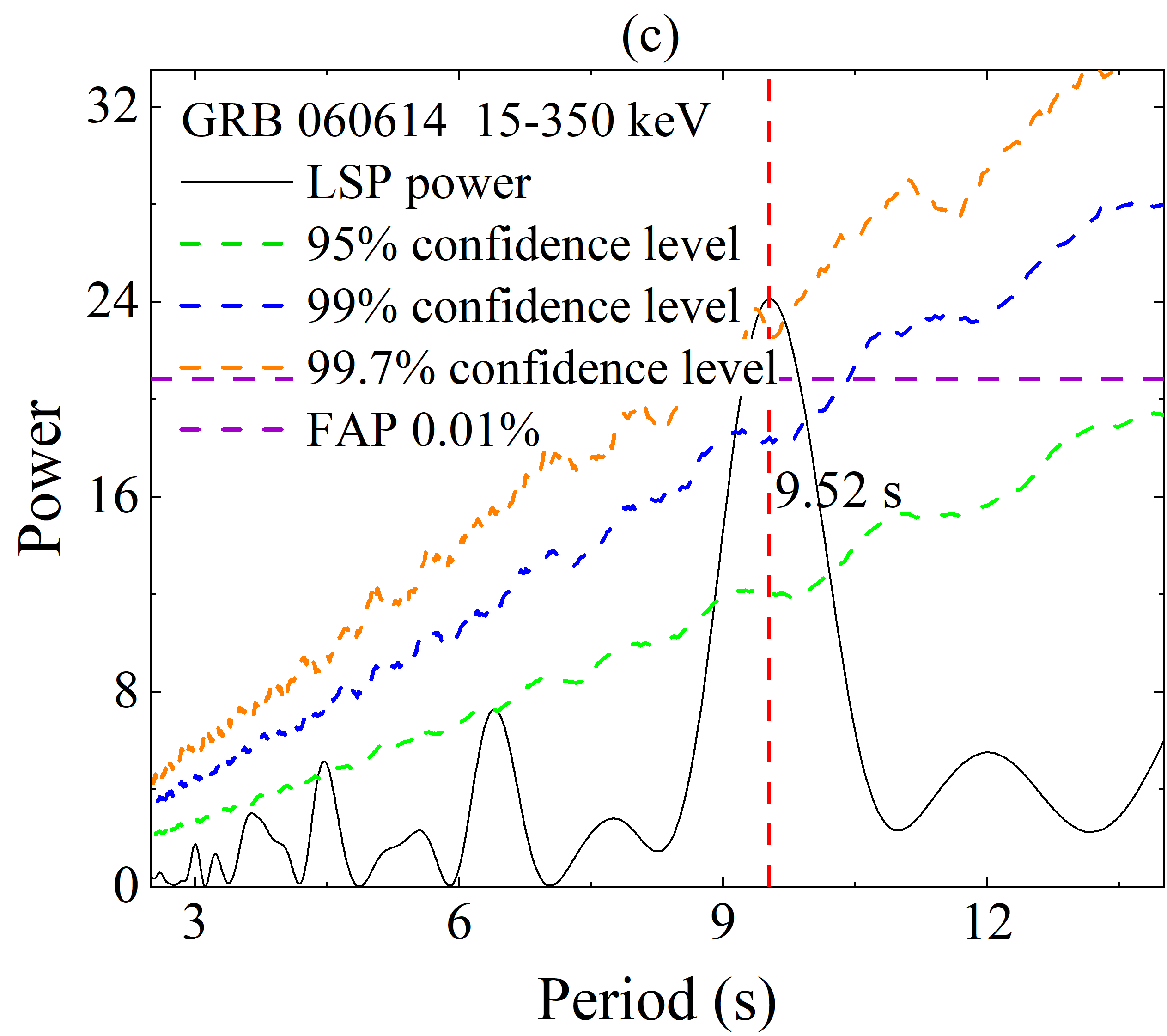}
\includegraphics[height=5.1cm,width=6.5cm]{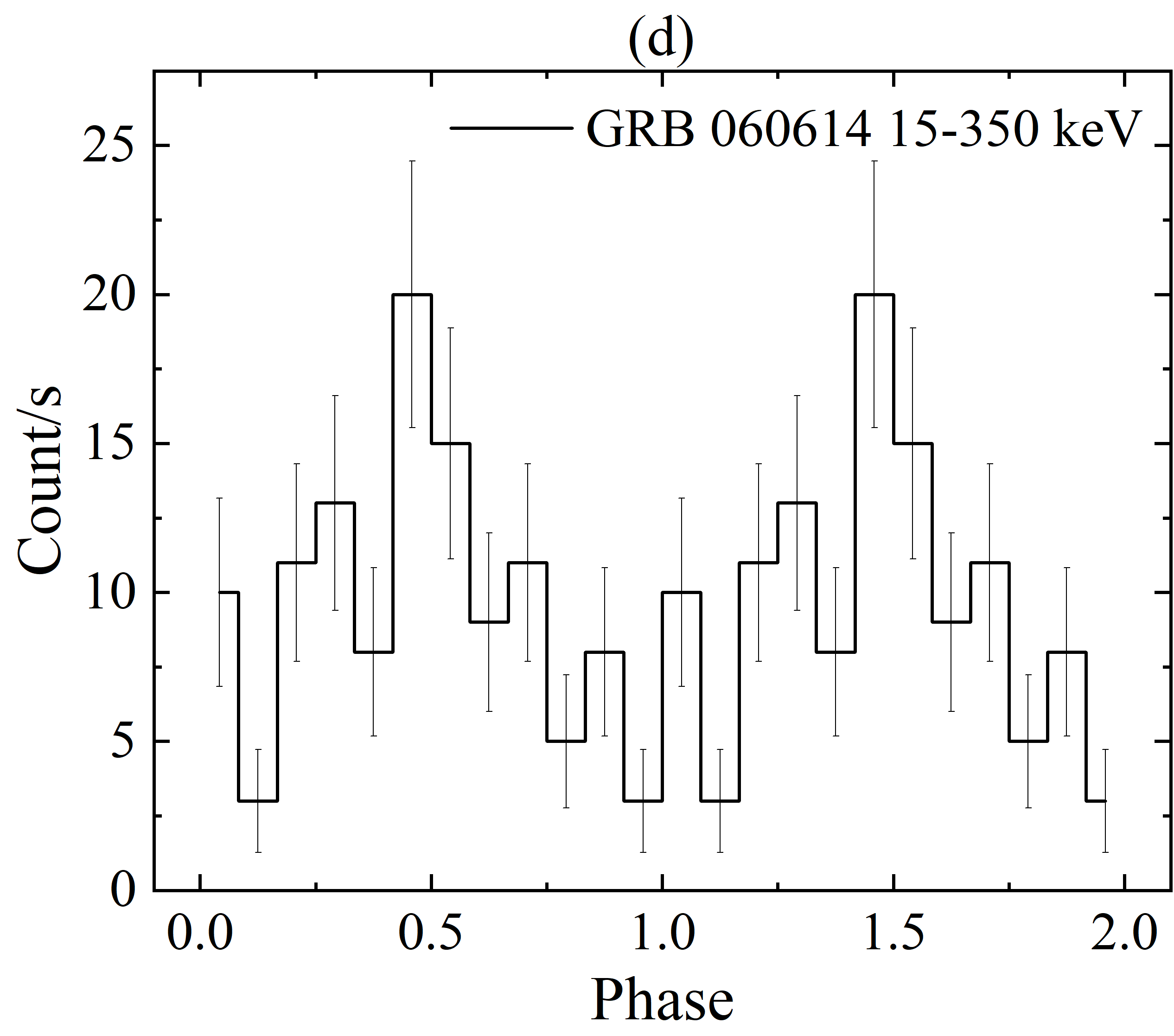}
\end{minipage}
\caption{Same as in Figure \ref{fig10} but for GRB 060614 (15-350 keV), note that the orange dashed line indicates the 99.7\% confidence level.}
\label{fig13}
\end{figure*}

\begin{figure*}
\begin{minipage}[t]{1\textwidth}
\centering
\includegraphics[height=5.2cm,width=8.7cm]{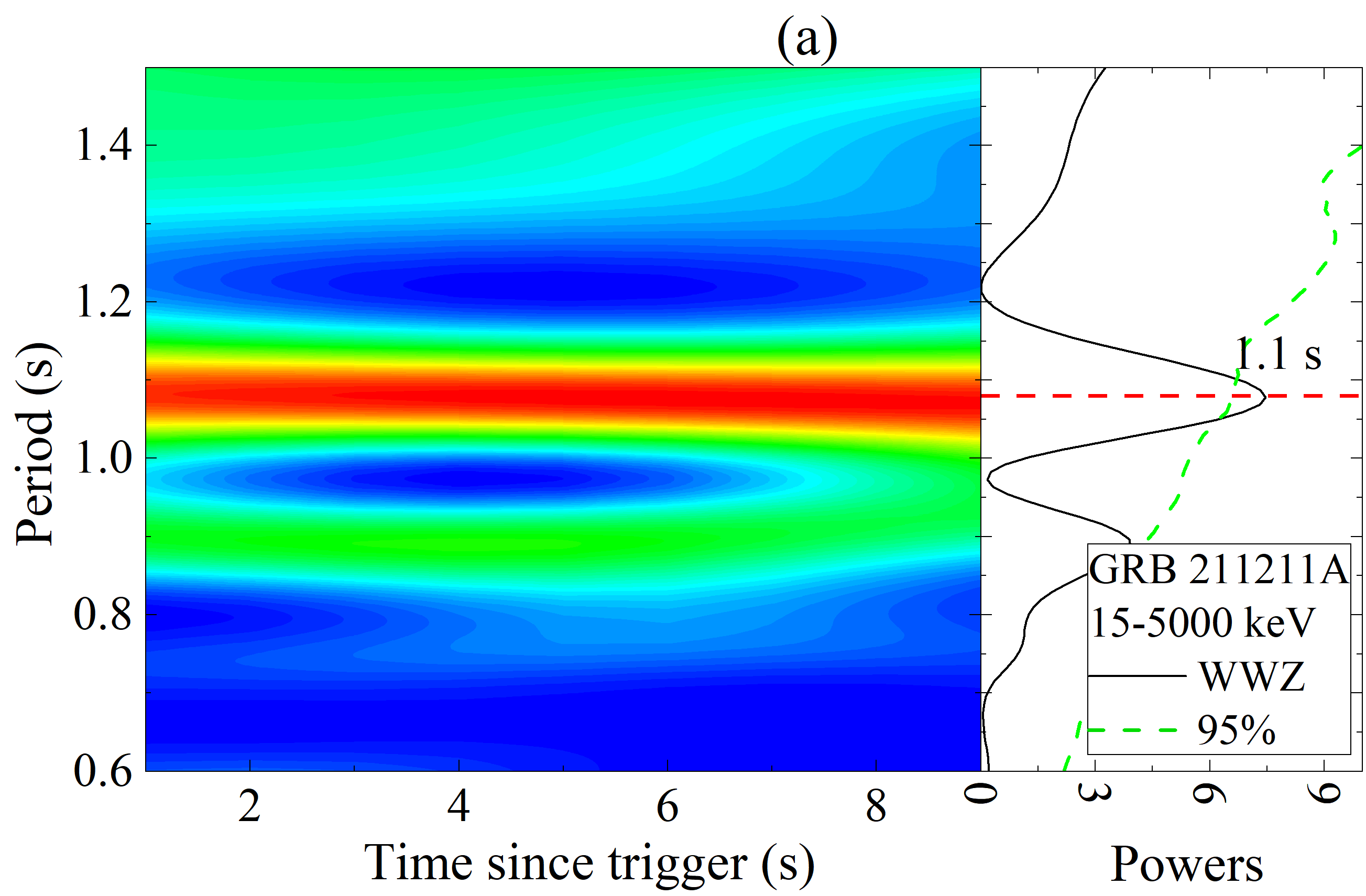}
\includegraphics[height=5.2cm,width=5.9cm]{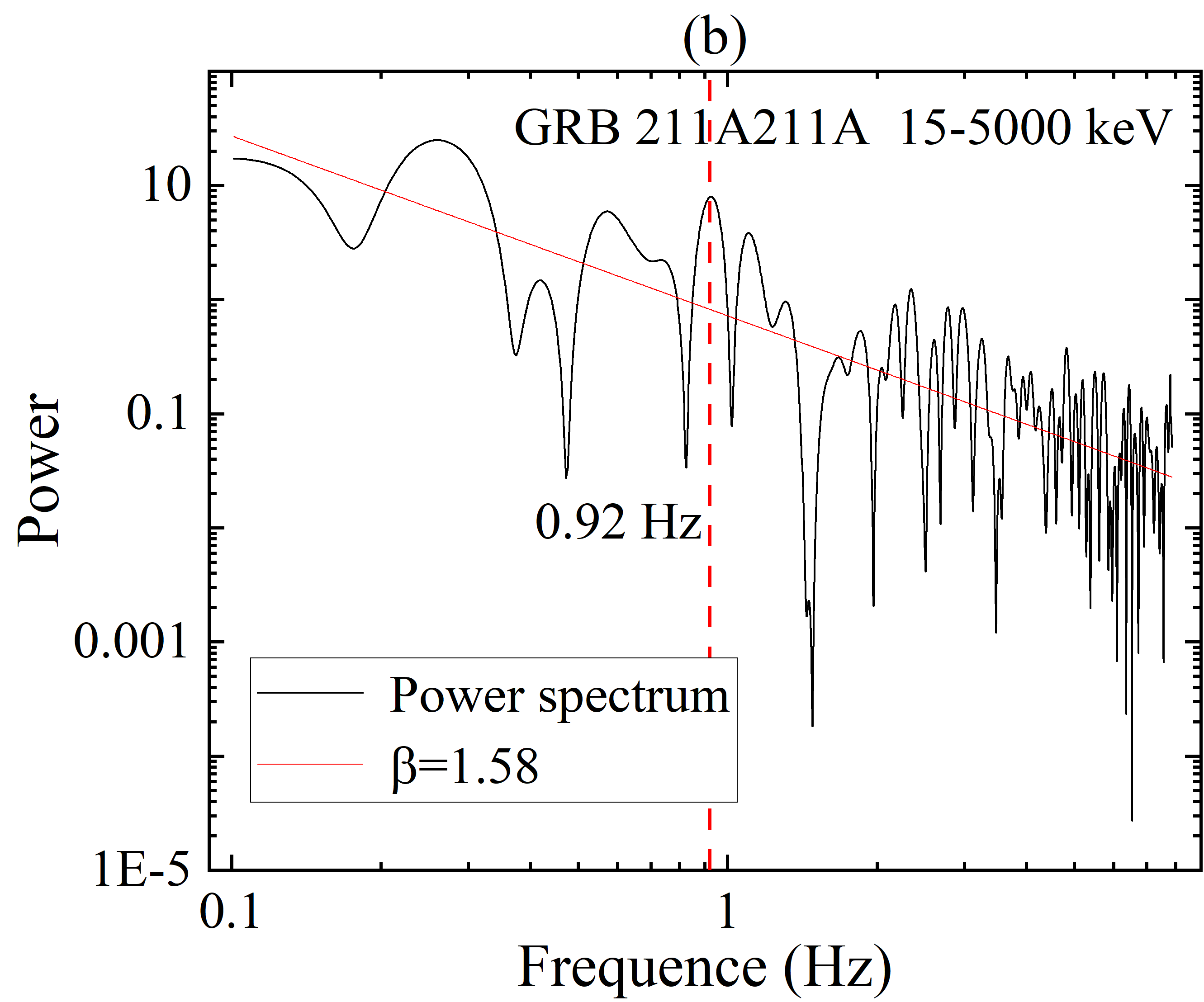}
\centering
\includegraphics[height=5.1cm,width=6.5cm]{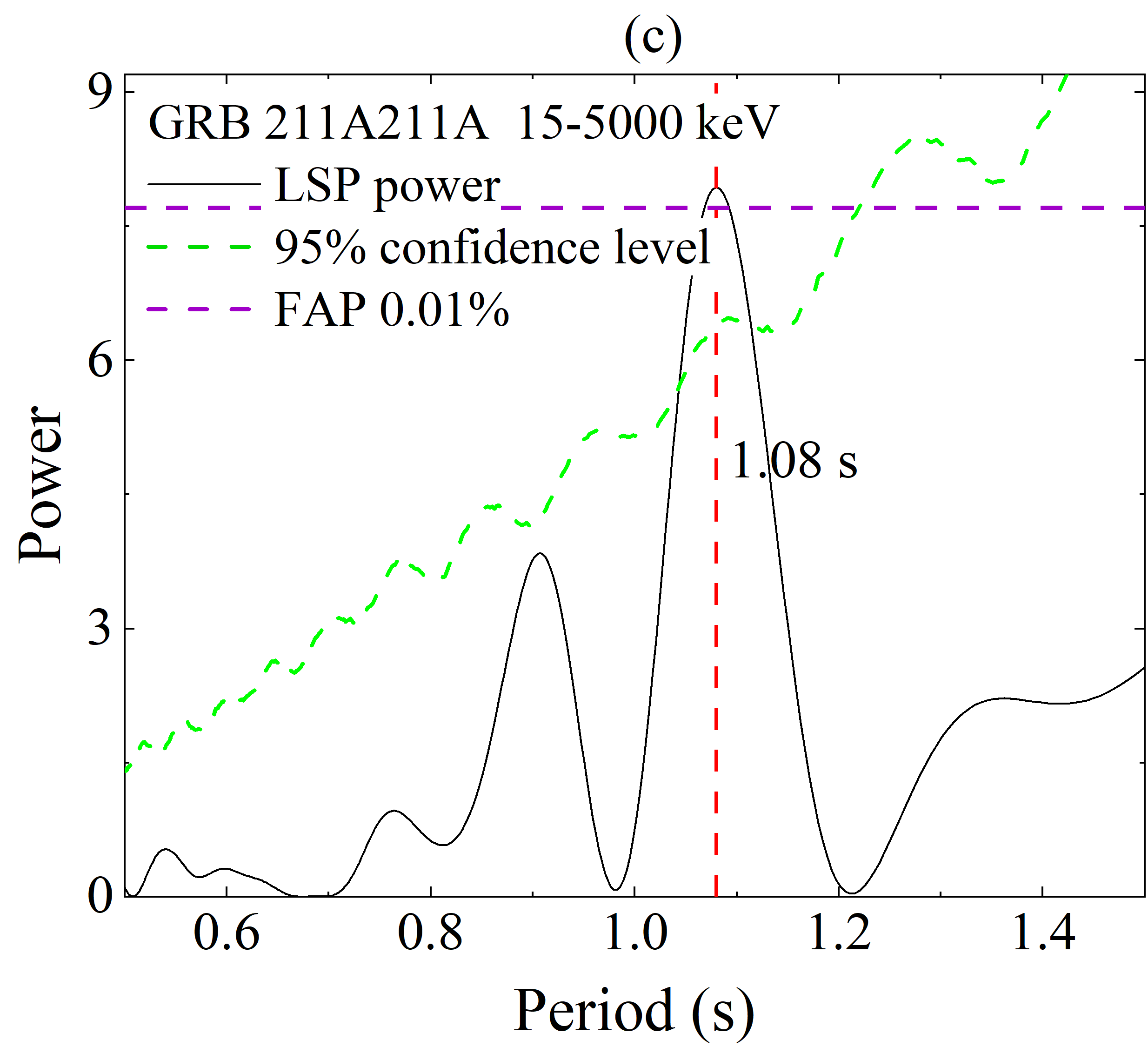}
\includegraphics[height=5.1cm,width=6.5cm]{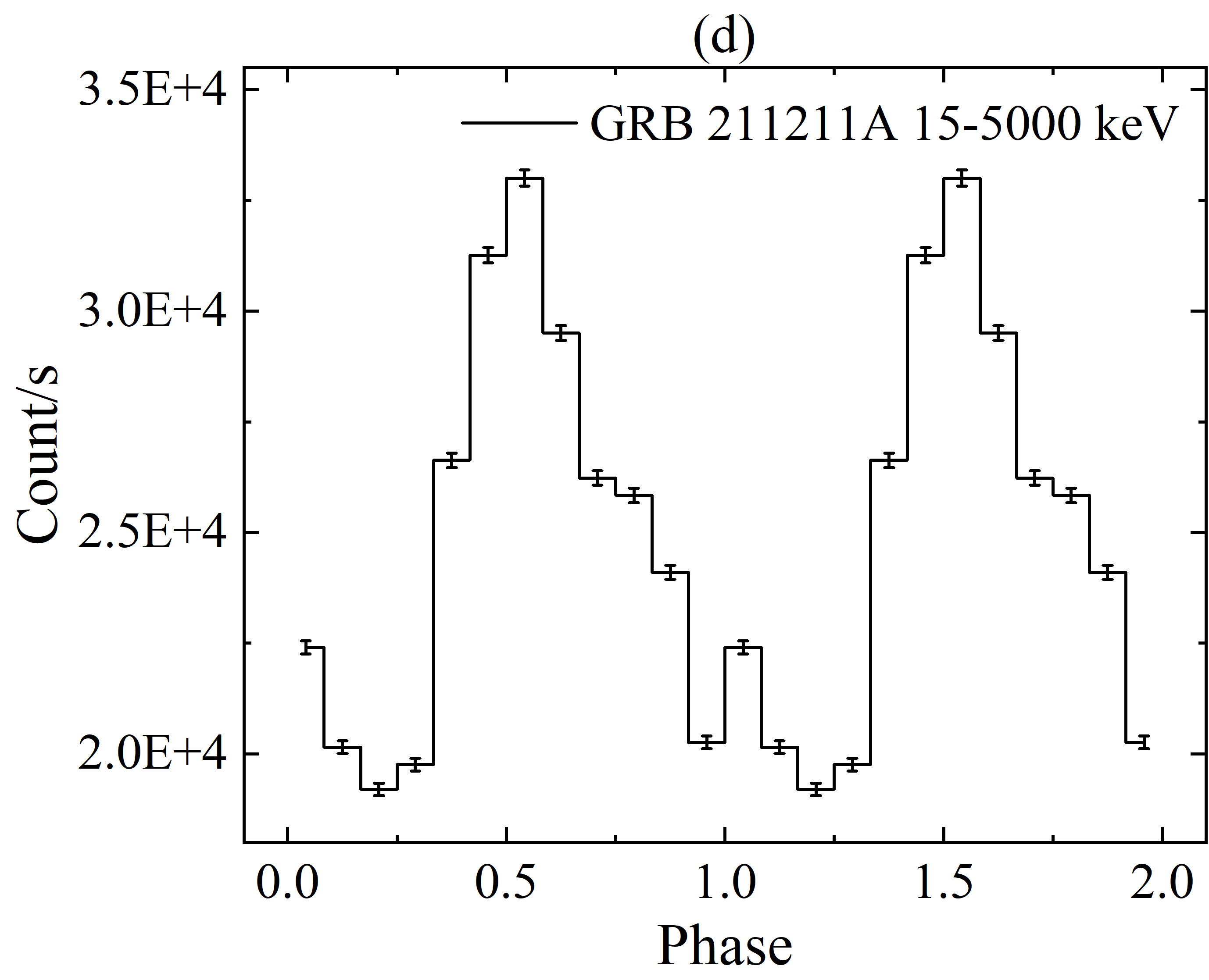}
\end{minipage}
\caption{Same as in Figure \ref{fig10} but for GRB 211211A (15-5000 keV).}
\label{fig14}
\end{figure*}

Similarly, the QPM analysis results for GRB 230307A in the energy range of 300-500 and 100-300 keV, as well as for GRBs 060614 and 211211A, are presented in Figures \ref{fig11}, \ref{fig12}, \ref{fig13}, and \ref{fig14}, respectively. In Table \ref{tabA1}, you can find the QPM results for all the light curves.

Indeed, although the periodicity within each light curve may not be immediately noticeable, the modulation process (QPM) in each light curve is discernible. It is intriguing to note that this QPM phenomenon simultaneously occurs in these three GRBs, which are believed to have a common origin. These QPMs may provide new clues for understanding the progenitor stars of this class of GRBs.

\end{document}